\documentclass{article}

\usepackage{amsmath}
\usepackage{arxiv}
\usepackage[utf8]{inputenc} % allow utf-8 input
\usepackage[T1]{fontenc}    % use 8-bit T1 fonts
\usepackage{hyperref}       % hyperlinks
\usepackage{url}            % simple URL typesetting
\usepackage{booktabs}       % professional-quality tables
\usepackage{amsfonts}       % blackboard math symbols
\usepackage{nicefrac}       % compact symbols for 1/2, etc.
\usepackage{microtype}      % microtypography
\usepackage{mathtools}
\usepackage{url}
\usepackage{float}
\usepackage{placeins}
\usepackage{setspace}
\usepackage{adjustbox}
\usepackage{lineno}
\usepackage{natbib}
\usepackage{bm}
\usepackage{upgreek}
\usepackage{textcomp}
\usepackage{multirow}
\makeatletter
\newif\ifnobrackets
\renewcommand\@cite[2]{\ifnobrackets\else[\fi{#1\if@tempswa , #2\fi}\ifnobrackets\else]\fi\nobracketsfalse}

\makeatother

\title{Improving Bayesian radiological profiling of waste drums using Dirichlet priors, Gaussian process priors, and hierarchical modeling}

\author{
   Eric Laloy\thanks{\normalsize{Corresponding Author, \texttt{eric.laloy@sckcen.be}}} \\
   \And
   Bart Rogiers \\
   \And
   An Bielen \\
   \And
   Alessandro Borella \\
   \And
   Sven Boden \\  
}

\begin{document}
\maketitle

\begin{abstract}
We present three methodological improvements of the ``SCK CEN approach" for Bayesian inference of the radionuclide inventory in radioactive waste drums, from radiological measurements. First we resort to the Dirichlet distribution for the prior distribution of the isotopic vector. The Dirichlet distribution possesses the attractive property that the elements of its vector samples sum up to 1. Second, we demonstrate that such Dirichlet priors can be incorporated within an hierarchical modeling of the prior uncertainty in the isotopic vector, when prior information about isotopic composition is available. Our used Bayesian hierarchical modeling framework makes use of this available information but also acknowledges its uncertainty by letting to a controlled extent the information content of the indirect measurement data (i.e., gamma and neutron counts) shape the actual prior distribution of the isotopic vector. Third, we propose to regularize the Bayesian inversion by using Gaussian process (GP) prior modeling when inferring 1D spatially-distributed quantities. As of uncertainty in the efficiencies, we keep using the same stylized drum modeling approach as proposed in our previous work to account for the source distribution uncertainty across the vertical direction of the drum. A series of synthetic tests followed by application to a real waste drum show that combining hierarchical modeling of the prior isotopic composition uncertainty together with GP prior modeling of the vertical Pu profile across the drum works well. We also find that our GP prior can handles both cases with and without spatial correlation. Of course, our GP prior modeling framework only makes sense in the context of spatial inference. Furthermore, the computational times involved by our proposed approach are on the order of a few hours, say about 2, to provide uncertainty estimates for all variables of interest in the considered inverse problem. This warrants further investigations to speed up the inference. 
\end{abstract}

\section{Introduction}
\label{intro}

Different variants of Bayesian data inversion or inference \citep[see, e.g.,][]{Gelman2014} have been recently introduced in the field of radiological characterization, independently by different groups \citep[][]{Carasco2021,Laloy2021,Buecherl2021,Clement2021}. All these approaches resort to Markov chain Monte Carlo (MCMC) sampling to derive the posterior activity or, equivalently, mass distribution of one or more radionuclides contained in the considered object, from radiological measurement data (e.g., gamma measurements). The ``SCK CEN approach" proposed by \citet{Laloy2021} has the following distinctive features: (i) it relies on an efficient MCMC algorithm called Hamiltonian Monte Carlo (HMC), (ii) accounts for two important sources of uncertainty, namely the measurement uncertainty and the uncertainty in the source distribution within the drum, and (iii) formulates an efficiency model that permits spatially-distributed inference along the vertical direction of the drum. In this work, we significantly improve upon this approach in the following four ways. 
\begin{itemize}
	\item First, we consider prior distributions that allow to work with a total Pu mass and associated isotopic vector directly. This permits accounting for more prior knowledge and results into a more realistic inference than when considering the activities of the Pu isotopes to be independent from each other. Our joint inference of total Pu mass and associated isotopic vector is done by putting a Dirichlet prior distribution on the isotopic vector.
	\item Second, we regularize the inferred spatial distribution of each considered (Pu-related) nuclide by using a Gaussian process (GP) prior for the spatially distributed total Pu mass \citep[e.g.,][]{Rasmussen-Williams2006}. The kernel parameters of the GP are then jointly inferred with the other variables. By regularizing the inverse solution, this GP prior largely improves computational tractability of the inference process.
	\item Third, we use Bayesian hierarchical modeling \citep[also called Bayesian multilevel modeling, see,][]{Gelman2014} to account for the uncertainty in the parameters of the Dirichlet prior that is put on the isotopic vector. Doing so allows the isotopic vector to vary spatially (here between horizontal drum segments), while still pooling information across the considered spatial locations (drum segments).
	\item Lastly, we combine two measurement techniques for a more accurate radiological characterization. Hence, we combine segmented gamma scanning (SGS) with passive neutron coincidence counting \citep[PNCC,][]{Borella2021}. This permits inference of both gamma emitting and non-gamma emitting nuclides, albeit not spatially for the non-gamma emitting nuclides. This because the PNCC technique provides a single measurement for the whole drum, of which the spatial information content is too low for resolving the nuclides' spatial distribution.
	
\end{itemize}
To the best of our knowledge, this is the first application of Dirichlet priors, Gaussian process priors and Bayesian hierarchical modeling, respectively, to Bayesian radiological characterization. Furthermore, the combination of gamma measurements with neutron coincidence counting (and other measurement techniques) within a Bayesian framework is a key task of the EU-funded CHANCE project and is new to this study and the companion CHANCE study of \citet{Carasco2022}.

The remainder of this paper is organized as follows. Section \ref{methods} describes the considered waste drum and the measurement data, while sections \ref{eff_models} and \ref{bayes} detail the various aspects of our MCMC-based Bayesian uncertainty quantification approach. Section \ref{results} then presents and discusses the obtained spatially distributed radionuclide inventory for both a synthetic and a real case, before section \ref{conclusion} summarizes our main findings and outlines some future developments.

\section{Measurements}
\label{methods}

\subsection{Segmented gamma scanning}
\label{sgs}

The segmented gamma scanning (SGS) system used in this work to quantify the gamma-emitters present in a radioactive waste drum is the same as the system used by \citet{Laloy2021}. For brevity we refer the reader to that study for a full description.

\subsection{ISOCS measurement}
\label{isocs_gamma}

An open geometry ISOCS gamma measurement with fixed detector position was performed \citep[][]{Boden2021} in addition to the 3AX-SGS measurement. In this work, this open geometry gamma measurement only serves to derive preliminary estimates of the isotopic vector components for the whole drum, by application of the FRAM isotopic composition software \citep{fram}. This software provides estimates of the relative proportions of gamma-emitting Pu isotopes based of their spectrum. It also calculates the relative proportions of $^{\rm 242}$Pu, $^{\rm 237}$Np and $^{\rm 235}$U based on empirical relationships. We would like to stress that our approach does not at all strictly require this open geometry measurement. Here we simply use it to derive a prior estimate for the isotopic vector but using any other source of information (e.g., expert knowledge, measured data from a relatively similar drum, ...) to derive this estimate would be just fine too.

\subsection{Passive neutron coincidence counting}
\label{ncc}

Passive neutron coincidence measurements (PNCC) were also carried out on  the considered drum. Generally, radioactive materials mainly emit neutrons according to three processes: $\left(\alpha,n\right)$ reactions, spontaneous fission, and neutron-induced fission. While  $\left(\alpha,n\right)$ reactions always emit only one neutron per event, more than one neutron may be emitted in a fission event. This results in a time correlation between detected events that is used in PNCC to estimate the quantity of material undergoing spontaneous fission in the sample being measured \citep{Dierckx-Hage1983}. In this study, PNCC measurements were carried out with a transportable system consisting of two identical slab counters equipped with $^3$He detectors each and coupled to shift register electronics. The system is fully described in \citet{Borella2021} and for brevity, we refer the reader to this publication for more information.

\subsection{Waste package and measurement setups}
\label{setup}

We focus on the real non-conditioned 200-liter radioactive waste drum considered in our previous publication on Bayesian radiological characterization \citep{Laloy2021}. This drum is fully described in \citet{Laloy2021} and here too we refer the reader to this publication for more details. Similarly as for our previous study, the drum dimensions, filling degree, matrix composition and matrix density distribution are assumed to be known and kept fixed. However, our approach accounts for the source distribution uncertainty. The source distribution uncertainty is a major source of uncertainty for this type of measurements.

\begin{figure}[hbt!]
	\noindent\hspace{1cm}\includegraphics[width=35pc]{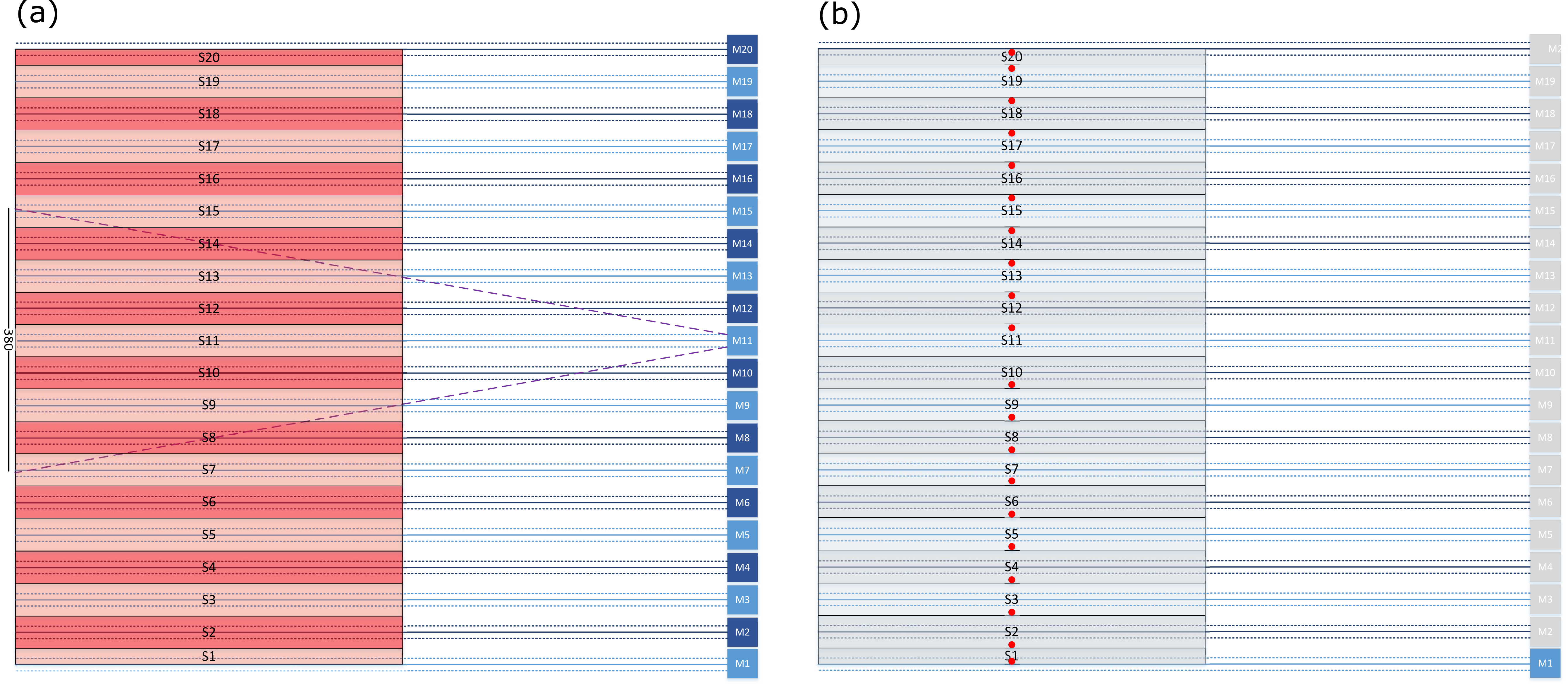}
	\caption{Vertical cross sections of the 3AX measuring setup for a 200-L drum. (a) 20 positions of the detector (M1 up to M20, blue) indicating the solid angle umbra (blue) for each measurement and the penumbra for measurement M11. The contents of the 200-L drum is discretized into 20 cylindrical volumes (segments) (red): S2 up to S19 in the middle of the drum with a height of 46.8 mm and S1 (bottom) \& S20 (top) with a height of 23.4 mm. (b) Homogeneous (gray rectangles) and point  (red dots) source distribution within each segment.}
	\label{fig1}
\end{figure}

As detailed in section \ref{eff_model_sgs}, the drum is discretized into 20 horizontal segments (Figure \ref{fig1}). The bottom and top segments (S1 \& S20) are 23.4 mm thick will the other 18 segments all have a thickness of 46.8 mm (S2 $-$ S19). To each segment corresponds a measurement with the detector directly in front of it. The 3AX-SGS measurement setup details are as follows:
\begin{itemize}
	\item The detector is surrounded by a rectangular lead slit collimator with a 1 mm copper coating (5$^{\circ}$ opening; 18 mm x 110 mm; 100 mm depth);
	\item The 200-liter drum is placed on a turning table and is continuously rotated during the measurements, with a rotation rate of 10 rotations per minute.
	\item The measured 20 segments result in 20 individual spectra (M1 up to M20) and measurement time is 300 s. For the first measurement, the bottom of the drum is placed in front of the middle of the detector (M1). For each following measurement the detector is raised by 46.8 mm (Figure \ref{fig1}). 
\end{itemize}

\FloatBarrier

Table \ref{table1} lists the identified nuclides in the drum  (independently by both the 3AX-SGS and open geometry ISOCS measurements) and, for the gamma-emitting ones, their associated photon energy peaks.
\begin{table}[hbt!]
	\caption{Studied nuclides and whenever applicable, associated photon energies, whether a FRAM-based prior estimate is available and the FRAM-based isotopic composition estimates together with their relative uncertainties (as provided by the FRAM software). Regarding the FRAM-derived isotopic composition, ``spectrum" signifies that the nuclides ' spectrum is used for the estimation while ``empirical" means application of an empirical relationship. Furthermore, $NA$ stands for ``not applicable" while COMP and RSD denote the FRAM-based estimate and its uncertainty in terms of relative standard deviation. The COMP values are percentages of the total Pu mass for the five Pu isotopes: $^{\rm 238}$Pu, $^{\rm 239}$Pu, $^{\rm 240}$Pu, $^{\rm 241}$Pu and $^{\rm 242}$Pu. As of $^{\rm 241}$Am, $^{\rm 237}$Np and $^{\rm 235}$U, the COMP values are the considered nuclide to total Pu mass ratios. The relatively low RSD values are associated with a very long measurement time of about 13 hours.}
	\begin{adjustbox}{center}
		\begin{tabular}{ccccc}%
			\hline
			Nuclide & Energy [keV] & FRAM & COMP & RSD [\%]\\
			\hline
			\multirow{3}{*}{$^{\rm 241}$Am} & 125.3 & \multirow{3}{*}{spectrum} & 0.043 [-] & 0.3 \\
			& 662.4 & & & \\
			& 722.01 & & & \\
			& & & & \\
			$^{\rm 244}$Cm & $NA$ & $NA$ & $NA$ & $NA$ \\
			& & & & \\
			$^{\rm 237}$Np & $NA$ & empirical & 0.002 [-] & 0.5 \\
			& & & & \\
			\multirow{2}{*}{$^{\rm 238}$Pu} & 152.72 & \multirow{2}{*}{spectrum} & 0.64 [\%] & 0.5\\
			& 766.39 & & & \\
			& & & & \\
			\multirow{5}{*}{$^{\rm 239}$Pu} & 129.30 & \multirow{5}{*}{spectrum} & 77.69 [\%] & 0.3 \\
			& 345.01 & & & \\
			& 375.05 & & & \\
			& 413.71 & & & \\
			& 451.48 & & & \\
			& & & & \\
			$^{\rm 240}$Pu & 160.31 & spectrum & 18.55 [\%] & 1.1 \\
			& & & & \\
			$^{\rm 241}$Pu & 148.57 & spectrum & 1.32 [\%] & 0.3 \\
			$^{\rm 242}$Pu & $NA$ & empirical & 1.80 [\%] & 1.1 \\
			$^{\rm 235}$U & $NA$ & empirical & 0.074 [-] & 1.9 \\
			\hline
		\end{tabular}
	\end{adjustbox}
	\label{table1}
\end{table}
\FloatBarrier

For the PNCC measurement, we consider the same drum discretization and treatment of the source distribution uncertainty as for the SGS measurement (Figure \ref{fig1}).

\section{Efficiency models}
\label{eff_models}

\subsection{Efficiency model of the SGS measurement}
\label{eff_model_sgs}

This subsection is slightly modified from our previous study \citep[][]{Laloy2021}. For our approach to account for the effect of the source distribution uncertainty within the drum on the detector efficiencies, it is required to model the whole measurement system for a limited set of prescribed configurations. To perform this modeling we used the complex cylinder model of the Geometry Composer V4.3 library of the ISOCS/LabSOCS software by \citet{Canberra2020}. The following two relatively extreme source distributions are considered (Figure \ref{fig1}):
\begin{itemize}
	\item Optimistic efficiency hypothesis: each segment contains an homogeneous source distribution (hypothesis $h$, Figure \ref{fig1}a). This is our chosen upper limit in efficiency for the considered fixed matrix properties.
	\item Conservative efficiency hypothesis: each segment contains a point source (hypothesis $p$) that is located on the drum axes on the border of a segment (on the bottom of the segment for S1 up to S10 and on the top of the segment for S11 up to S20, see Figure \ref{fig1}b). This is the minimum efficiency for the considered fixed matrix properties.
\end{itemize}
The drum dimensions, filling degree, mean matrix composition and density remain constant and fixed in the ISOCS model. In contrast, the source distribution is varied. In case an homogeneous source distribution is modeled, the matrix shielding is minimal for the part of the source close to the detector and increases more towards the center of the drum. For the modeling of a point source located on the drum axes, the matrix shielding thickness is larger than the drum radius. This because the detector points to the center of a sub-volume but in this case the point source is located at the bottom or top of the sub-volume (Figure \ref{fig1}b). The symmetry between measurements M1-M10 and measurements M11-M20 (Figure \ref{fig1}) allows for duplicating the efficiency calculations for the lower part of the drum to the upper part of the drum, therebey reducing the required amount of calculations by two. An individual model is thus designed for every individual measurement associated with the lower part of the drum (M1 up to M10), with the source located stepwise in segments S1 up to S20 for both the homogeneous (hypothesis $h$) and point source (hypothesis $p$) scenarios, respectively. This leads to a total of 400 models (10 detector locations $\times$ 20 source locations $\times$ 2 source distributions). For both the $h$ and $p$ assumptions, the detector's efficiency is therefore computed for each configuration of source location (S, the source is located in a given segment while the other 19 segments do not contain any source), detector location (M), and photon energy (among the 12 energies listed in Table \ref{table1}). These efficiency values are then used to create two $n_{source}$ $\times$ $n_{peaks}$ $\times$ $n_{seg}$ 3D arrays, $\textbf{E}_{\rm SGS}^h$ and $\textbf{E}_{\rm SGS}^p$, for the $h$ and $p$ assumptions, respectively. Here $n_{source} = n_{seg} = 20$ and $n_{peaks} = 12$. Note that an even more favorable or ``positively" extreme distribution than hypothesis $h$ consists of a circularly distributed source at a radius close to the drum's radius. Also our $h$ and $p$ assumptions are relatively extreme source distributions for the considered (fixed) matrix properties only. Other matrices, such as one including one or more self-shielding sources such as fuel pellets for instance, could lead to other extreme situations \citep[see also][for other examples]{Martens-Filss1990}. It is worth noting that, in principle, our approach allows for including not just two ($h$ and $p$) but a few ``end-member" (that is,``extreme" case) efficiencies that are considered relevant. This will be explored in future work.

The efficiency $\epsilon\left(e\right)$ corresponding to a given gamma detection at energy $e$ can be balanced between the $h$ and $p$ assumptions for a given segment by using a $\lambda \in \left[0,1\right]$ coefficient: $\epsilon\left(e\right) = \lambda\epsilon^h\left(e\right) + \left(1 - \lambda\right)\epsilon^p\left(e\right)$. This forms the basis of our count simulation model. For a given count rate, $c^{rate}$, we have

\begin{linenomath}
	\begin{equation}
	\label{3ax1}
	c^{rate}_{i,k} = \sum\limits_{l=1}^{n_{seg}}a_{l}^{j}P_{\gamma_{j,k}}\left[\lambda_l\epsilon^h_{i,l}\left(e_k\right)+\left(1 - \lambda_l\right)\epsilon^p_{i,l}\left(e_k\right)\right],
	\end{equation}
\end{linenomath}
where the subscript $i$ indexes the detector location, $i=1, \cdots, 20$, the subscript $k$ indexes the considered energy peak,  $k=1,\cdots,12$, the superscript $j$ indexes the considered nuclide, $j=1, \cdots, 5$, that is responsible for peak $k$, $a_{l}^{j}$ denotes the activity of nuclide $j$ at segment location $l$, and $P_{\gamma_{j,k}}$ is the emission probability of nuclide $j$ at energy $k$.

The simulated gross count at detector location $i$ for the energy $k$, $c^{gross}_{i,k}$, is then obtained from
\begin{linenomath}
	\begin{equation}
	\label{3ax2}
	c^{gross}_{i,k} = c^{rate}_{i,k}t_m+b_{i,k},
	\end{equation}
\end{linenomath}
where $t_m$ is the measurement time and $b_{i,k}$ is the background continuum count for time $t_m$ at detector location $i$ for the energy $k$.

\subsection{Efficiency model of the PNCC measurement}
\label{eff_model_ncc}

As stated earlier, the PNCC setup and simulation model used herein are detailed in \citet{Borella2021} and a full mathematical development is given in \citet{Borella-Rossa2022}. We only describe here what is needed to understand our specific inversion approach. 

Similarly as for the 3AX-SGS setup, PNCC efficiencies for the different source configurations shown in Figures \ref{fig1}a - b were simulated, thereby leading the creation of two 20-dimensional vectors, $\textbf{e}_{\rm PNCC}^h$ and  $\textbf{e}_{\rm PNCC}^p$. The spontaneous fission ($SF$) decay of $^{\rm 240}$Pu was simulated in \citet{mcnp} to determine the efficiencies, here defined as number of reals per spontaneous fission event. The modeled observable is the number of reals per second, $R_s$, and in this work is expressed as

\begin{linenomath}
	\begin{equation}
	\begin{split}
	\label{ncc1}
	R_s & = Fs_{240Pu}\sum\limits_{l=1}^{n_{seg}}m_{240Pu_{eq},l}\left[\lambda_l\epsilon^h_{\rm{PNCC},l}+\left(1 - \lambda_l\right)\epsilon^p_{\rm{PNCC},l}\right], \\
	m_{240Pu_{eq},l} & = m^{all}_{l}\sum\limits_{i=1}^{n_{n}}\displaystyle\frac{Q_i}{Q_{240Pu}}m^f_{i,l}, \\
	\ Q_i & = Fs_i\nu_{s(2),i},
	\end{split}
	\end{equation}
\end{linenomath}

where $m_{240Pu_{eq}}$ represents the mass of $^{\rm 240}$Pu that results in the same $Rs$ as the considered sample, $Fs_i$ is the spontaneous fission rate per mass unit of radionuclide $i$ (negligibly small for odd Pu isotopes, $^{\rm 237}$Np and $^{\rm 235}$U), $\epsilon^h_{\rm{PNCC},l}$ and $\epsilon^p_{\rm{PNCC},l}$ are the pre-calculated efficiencies for the homogeneous and point source distribution assumptions, respectively, in segment $l$, $m^{all}_{l}$ is the inferred total mass of the considered $n_{n}$ radionuclides in segment $l$, $m^f_{i,l}$ is the inferred mass fraction of radionuclide $i$ in segment $l$, and $\nu_{s(2),i}$ is the second factorial moment of the neutron emission distribution by spontaneous fission associated to radionuclide $i$.

\section{Bayesian inference}
\label{bayes}

\subsection{Bayesian paradigm}

The forward problem is commonly represented as
\begin{equation}
\textbf{d} = F\left(\bm{\uptheta}\right) + \bm{\upeta},
\label{mcmc0}
\end{equation}

where $\textbf{d} = \left[d_1, \ldots, d_{N_d} \right] \in \mathbb{R}^{N_d}, N_d \geq 1$ is the measurement data, $F\left(\bm{\uptheta}\right)$ is a deterministic forward model with parameters $\bm{\uptheta}$ and the noise term $\bm{\upeta}$ lumps all sources of errors. 

In the Bayesian paradigm, parameters in $\bm{\uptheta}$ are viewed as random variables with a posterior probability density function (pdf), $p\left(\bm{\uptheta} | \textbf{d} \right)$, given by
\begin{linenomath*}
	\begin{equation}
	p\left(\bm{\uptheta} | \textbf{d}  \right) = \frac{p \left(\bm{\uptheta}\right) p \left(\textbf{d} | \bm{\uptheta}\right)}{p \left( \textbf{d} \right)} \propto p\left(\bm{\uptheta}\right) p\left(\textbf{d} | \bm{\uptheta}\right),
	\label{mcmc1}
	\end{equation}
\end{linenomath*}
where $p \left(\textbf{d} | \bm{\uptheta}\right)$ signifies the likelihood function of $\textbf{d}$ given $\bm{\uptheta}$. The normalization factor $p \left( \textbf{d} \right) = \int  p\left(\bm{\uptheta}\right) p\left(\textbf{d} | \bm{\uptheta}\right) d\bm{\uptheta}$ is not needed for parameter inference when the parameter dimensionality is kept fixed. In the remainder of this paper, we will thus focus on the unnormalized density $p\left(\bm{\uptheta} | \textbf{d} \right) \propto p\left(\bm{\uptheta}\right) p\left(\textbf{d} | \bm{\uptheta}\right)$. 

\subsection{Inferred variables}
\label{inf_var}
Here the vector of inferred variables, $\bm{\uptheta}$, consists of the following

\begin{enumerate}
	
	\item The isotopic vector (\%) of the following five Pu isotopes, $^{\rm 238}$Pu, $^{\rm 239}$Pu, $^{\rm 240}$Pu, $^{\rm 241}$Pu and $^{\rm 242}$Pu, in each of the 20 segments
	\begin{linenomath*}
		\begin{equation}
		\begin{split}
		\textbf{v}_i & = \left[m^f_{{\rm 238Pu}, i}, m^f_{{\rm239Pu}, i}, m^f_{{\rm 240Pu}, i}, m^f_{{\rm 241Pu}, i}, m^f_{{\rm 242Pu}, i} \right] \\
		\rm{with} \ i & = 1, \cdots, 20, \rm{the \ segment \ number}.
		\label{vvec}
		\end{split}
		\end{equation}
	\end{linenomath*}
	For notational convenience, we encapsulate the 20 isotopic vectors in the 2D \textbf{V} array
	\begin{linenomath*}
		\begin{equation}
		\textbf{V} = \begin{pmatrix} \textbf{v}_1 \\ \vdots \\ \textbf{v}_{20}\\ \end{pmatrix}.
		\label{Vmat}
		\end{equation}
	\end{linenomath*}
	\item The total Pu mass (g) in each of 20 segments, $\textbf{p}  = \left[p_1, \cdots, p_{20}\right]$.
	\item The $^{\rm 241}$Am mass (g), expressed as the ratio of the $^{\rm 241}$Am mass to the total Pu mass, in each of the 20 segments, $\textbf{a}  = \left[a_1, \cdots, a_{20}\right]$.
	\item The (unique) $cr$ parameter that is used to derive the $^{\rm 244}$Cm mass (g), $c$, in each of the 20 segments, $\textbf{c}  = \left[c_1, \cdots, c_{20}\right]$. As detailed in section \ref{cm244_prior}, the $cr$ parameter is equal to $\log_{10} \left[\left(m_{{\rm 239Pu},i}+m_{{\rm 240Pu},i}\right)/m_{{\rm 244Cm},i}\right]$ where $m_{{\rm 239Pu},i}$, $m_{{\rm 240Pu},i}$ and $m_{{\rm 244Cm},i}$ are the $^{\rm 239}$Pu, $^{\rm 240}$Pu and $^{\rm 244}$Cm masses in a given segment $i$.
	\item The $^{\rm 237}$Np mass (g), expressed as the ratio of the $^{\rm 237}$Np mass to the total Pu mass, in each of the 20 segments, $\textbf{n}  = \left[n_1, \cdots, n_{20}\right]$.
	\item The $^{\rm 235}$U mass (g), expressed as the ratio of the $^{\rm 235}$U mass to the total Pu mass, in each of the 20 segments, $\textbf{u}  = \left[u_1, \cdots, u_{20}\right]$.
	\item The $\bm{\uplambda} = \left[\lambda_1, \cdots, \lambda_{20}\right]$ vector.
	\item The 240 background continuum counts of the 3AX measurement, $\textbf{b}  = \left[b_1, \cdots, b_{240}\right]$ (12 energy peaks $\times$  20 detector locations).
	\item The five parameters, $\bm{\upalpha} = \left[\alpha_1, \cdots, \alpha_5\right]$, of the Dirichlet prior distribution  put on $\textbf{v}_1, \cdots, \textbf{v}_{20}$, when Bayesian hierarchical modeling is used to account for the uncertainty in our preliminary estimate of the Pu isotopic vector. Here, $\alpha_1$, $\alpha_2$, $\alpha_3$, $\alpha_4$, and $\alpha_5$ therefore represent the prior mass percentages of $^{\rm 238}$Pu, $^{\rm 239}$Pu, $^{\rm 240}$Pu, $^{\rm 241}$Pu and $^{\rm 242}$Pu, respectively. Notice that although they share the same prior, the $\textbf{v}_1, \cdots, \textbf{v}_{20}$ vectors are free to take different values if needed. In other words, the posterior isotopic vectors can be different between the 20 segments.
	\item When a GP prior is put on $\textbf{p}$ (see later on), the mean, $\mu_{\rm GP}$, variance, $\sigma^2_{\rm GP}$ and lengthscale, $\tau_{\rm GP}$, of the chosen GP kernel.
	
\end{enumerate}

Overall, we therefore jointly infer between 441 and 449 parameters depending on the considered setup.

Our measurement vector, $\textbf{d}$, contains the 240 measured gross counts by 3AX-SGS, $\textbf{d}_{SGS}$ and the number of reals per second measured by PNCC, $d_{\rm PNCC}$, $\textbf{d} = \left[\textbf{d}_{SGS}, d_{\rm PNCC}\right]$. In addition, our derived posterior distribution is not only conditioned to $\textbf{d}$ but also to the predefined $\textbf{E}_{\rm SGS}^h$, $\textbf{E}_{\rm SGS}^p$, $\textbf{e}_{\rm PNCC}^h$ and $\textbf{e}_{ \rm PNCC}^p$ arrays. This results in assessment of
\begin{linenomath*}
	\begin{equation}
	p\left(\bm{\uptheta}  | \textbf{d}, \textbf{E}_{\rm SGS}^h,\textbf{E}_{\rm SGS}^p, \textbf{e}_{\rm PNCC}^h,\textbf{e}_{\rm PNCC}^p\right) \propto p\left(\textbf{d}, \textbf{E}_{\rm SGS}^h,\textbf{E}_{\rm SGS}^p, \textbf{e}_{\rm PNCC}^h,\textbf{e}_{\rm PNCC}^p\right | \bm{\uptheta})p\left(\bm{\uptheta}\right).
	\label{mcmc2}
	\end{equation}
\end{linenomath*}

\subsection{Likelihood function}

Our likelihood function, $p\left(\textbf{d}, \textbf{E}_{\rm SGS}^h,\textbf{E}_{\rm SGS}^p, \textbf{e}_{\rm PNCC}^h,\textbf{e}_{\rm PNCC}^p | \bm{\uptheta}\right)$ is the product of the likelihoods of the SGS and PNCC data, respectively, given $\bm{\uptheta}$
\begin{linenomath*}
	\begin{equation}
	p\left(\textbf{d}, \textbf{E}_{\rm SGS}^h,\textbf{E}_{\rm SGS}^p, \textbf{e}_{\rm PNCC}^h,\textbf{e}_{\rm PNCC}^p | \bm{\uptheta}\right) = p\left(\textbf{d}_{\rm SGS}, \textbf{E}_{\rm SGS}^h,\textbf{E}_{\rm SGS}^p | \bm{\uptheta} \right) \times p\left(\textbf{d}_{\rm PNCC}, \textbf{e}_{\rm PNCC}^h,\textbf{e}_{\rm PNCC}^p | \bm{\uptheta} \right).
	\label{mcmc3}
	\end{equation}
\end{linenomath*}

If we assume $\textbf{d}_{SGS}$ to follow a Poisson process, which is the norm for count data, we have
\begin{linenomath*}
	\begin{equation}
	p\left(\textbf{d}_{\rm SGS}, \textbf{E}_{\rm SGS}^h,\textbf{E}_{\rm SGS}^p | \bm{\uptheta} \right) = \prod\limits_{i=1}^{N_{\rm SGS}}\exp\left(-\tilde{d_i}\right)\displaystyle\frac{\tilde{d_i}^{d_i}}{d_i!},
	\label{mcmc4}
	\end{equation}
\end{linenomath*}
where $\tilde{\textbf{d}}_{\rm SGS} = \left[\tilde{d}_1, \ldots, \tilde{d}_{N_{\rm SGS}} \right] = F_{\rm SGS}\left(\bm{\uptheta}\right)$ contains the simulated gross counts for a given $\bm{\uptheta}$.

For $d_{PNCC}$, we assume a normal distribution with a 5\% relative error, $\sigma_{\rm PNCC} = 0.05*\tilde{d}_{PNCC}$
\begin{linenomath*}
	\begin{equation}
	p\left(\textbf{d}_{\rm PNCC}, \textbf{e}_{\rm PNCC}^h,\textbf{e}_{\rm PNCC}^p | \bm{\uptheta} \right) = N\left(\tilde{d}_{PNCC}, 0.05\tilde{d}_{PNCC}\right),
	\label{mcmc5}
	\end{equation}
\end{linenomath*}
where $\tilde{d}_{PNCC}$ is the simulated number of reals per second corresponding to a given $\bm{\uptheta}$.

\subsection{Prior distributions}

\subsubsection{Dirichlet prior and hierarchical modeling of the prior uncertainty in the vector} 
\label{alpha_prior}
With respect to the Pu isotopic vector associated with each segment, that is, each row $\textbf{v}_i$ of $\textbf{V}$ with $i = 1, \cdots, 20$ and  $\textbf{v}_i$ = $\left[m^f_{{\rm 238Pu},i}, m^f_{{\rm 239Pu},i}, m^f_{{\rm 240Pu},i}, m^f_{{\rm 241Pu},i}, m^f_{{\rm 242Pu},i}\right]$, we use a Dirichlet prior distribution, $\textbf{v}$ $\propto$ $Dir\left(\bm{\upalpha}\right)$. This multidimensional distribution has the nice property that the values of the elements of any sampled vector $\textbf{v}$ sum up to one

\begin{linenomath*}
	\begin{equation}
	\begin{split}
	p\left(\textbf{v}\right) & = \displaystyle\frac{1}{\rm B \left(\bm{\upalpha}\right)}\prod\limits_{j=1}^{K}v_j^{\alpha_j-1}, \\
	{\rm where}\  \rm B \left(\bm{\upalpha}\right) & = \displaystyle\frac{\prod\limits_{j=1}^{K}\Upgamma\left(\alpha_j\right)}{\Upgamma\left(\sum\limits_{j=1}^{K}\alpha_j\right)} \ {\rm and}\  \bm{\upalpha} = \left[\alpha_1, \cdots, \alpha_K\right], \\
	{\rm with} \displaystyle \sum\limits_{j=1}^{K} v_j & = 1\ {\rm and}\  v_j \geq 0 \ \ \forall j \in \left\{1, \cdots, K\right\}.
	\label{mcmc6}
	\end{split}
	\end{equation}
\end{linenomath*}

In case there is no additional prior information on the 20 $\textbf{v}_i$ vectors, we set $\alpha_1$ = $\alpha_2$ = $\alpha_3$ = $\alpha_4$ = $\alpha_5$ = 1 in $\bm{\upalpha}$, which amounts to a flat Dirichlet prior where for each segment, each considered Pu isotope receives the same prior probability under the constraint that the Pu isotope fractions sum up to 1. 

Besides a flat Dirichlet prior, we also consider the case where prior information about the Pu isotopic vector is available. As written earlier, here this information comes from isotopic composition analysis performed using the FRAM software on the basis of an open geometry and fixed detector ISOCS measurement \citep[][]{Boden2021}, together with empirical relationships. Nevertheless, we would like to emphasize again that such measurement is not a requirement of our approach as the latter is suited to the general case when a prior vector estimate is available, no matter how it is obtained. When such estimate exists, we resort to Bayesian hierarchical modeling to jointly infer with the other unknowns the $\bm{\upalpha}$ vector of the Dirichlet prior distribution put on each of the $\textbf{v}_1, \cdots \textbf{v}_{20}$ vectors. This amounts to set $p\left(\textbf{v}\right)$ as

\begin{linenomath*}
	\begin{equation}
	p\left(\textbf{v}\right) = \int p\left(\textbf{v} | \bm{\upalpha} \right) p\left(\bm{\upalpha}\right)d\bm{\upalpha}.
	\label{mcmc7}
	\end{equation}
\end{linenomath*}

As of $p\left(\bm{\upalpha}\right)$, we put truncated normal priors on $\alpha_1, \cdots, \alpha_5$ with means equal to our preliminary estimates obtained from prior isotopic composition analysis, $\bm{\upmu}_{\alpha} = \left[\mu_{\alpha, 1}, \cdots, \mu_{\alpha, 5}\right]$ = $\left[\mu_{\alpha,{\rm 238Pu}}, \mu_{\alpha, {\rm 239Pu}}, \mu_{\alpha, {\rm 240Pu}}, \mu_{\alpha, {\rm 241Pu}}, \mu_{\alpha, {\rm 242Pu}}\right]$, (see section \ref{isocs_gamma}) and standard deviations set to half the associated means, $\sigma_{\alpha,i}  = 0.5\mu_{\alpha, i}$, which corresponds to a coefficient of variation of 50 \%

\begin{linenomath*}
	\begin{equation}
	\begin{split}
	p\left(\bm{\upalpha}\right) & = 
	\begin{cases}
	N\left(\bm{\upmu}_{\alpha}, \textbf{C}_{\alpha}\right), & \text{if all}\ \alpha_i > 0 \\ 
	0, & \text{otherwise}
	\end{cases}
	\\
	\textbf{C}_{\alpha} & = \begin{pmatrix} 
	\sigma_{\alpha,1}^2 & \cdots & 0 \\
	\vdots & \ddots & \vdots \\
	0 & \cdots & \sigma_{\alpha,5}^2 \\
	\end{pmatrix}.
	\label{mcmc8}
	\end{split}
	\end{equation}
\end{linenomath*}
In this study, the FRAM-based prior isotopic composition analysis gives: $m^f_{{\rm 238Pu}}$ = 0.0064, $m^f_{{\rm 239Pu}}$ = 0.7769, $m^f_{{\rm 240Pu}}$ = 0.1855, $m^f_{{\rm 241Pu}}$ = 0.0132 and $m^f_{{\rm 242Pu}}$ = 0.0180. The degree of enforcement of the prior isotopic vector is controlled by the width of the Dirichlet prior that is put on each $\textbf{v}_i$, which is itself defined by the actual $\alpha$ values. For $\alpha$ values $>$ 1, the larger the value the more the probability mass is concentrates near the  predefined proportions. In this work we set $\mu_{\alpha,{\rm 238Pu}}$ = 0.64, $\mu_{\alpha,{\rm 239Pu}}$ = 77.69, $\mu_{\alpha,{\rm 240Pu}}$ = 18.55, $\mu_{\alpha,{\rm 241Pu}}$ = 1.32 and $\mu_{\alpha,{\rm 242Pu}}$ = 1.80. This is deemed to feed the inference with a good level of prior knowledge with still allowing for enough deviations if required to fit the data.

\subsubsection{Gaussian process prior for the Pu mass profile}
\label{gp_prior}

Gaussian processes (GPs) form flexible nonparametric priors that are widely used for various machine learning tasks such as regression and classification \citep[see the book by][for details]{Rasmussen-Williams2006}. A GP is a collection of random variables, any finite number of which have a joint Gaussian distribution. A GP, $f\left(\textbf{x}\right)$, is completely determined by its mean function, $\mu\left(\textbf{x}\right)$, and covariance function, $k\left(\textbf{x},\textbf{x}'\right)$ 

\begin{linenomath*}
	\begin{equation}
	\begin{split}
	f\left(\textbf{x}\right) & \sim \mathcal{GP} \left(\mu\left(\textbf{x}\right),k\left(\textbf{x},\textbf{x}'\right)\right), \\
	\mu\left(\textbf{x}\right) & = \mathbb{E}\left[f\left(\textbf{x}\right)\right], \\
	k\left(\textbf{x},\textbf{x}'\right) & = \mathbb{E}\left[\left(f\left(\textbf{x}\right) - \mu\left(\textbf{x}\right)\right)\left(f\left(\textbf{x}'\right) - \mu\left(\textbf{x}'\right)\right)\right]. \\
	\label{mcmc9}
	\end{split}
	\end{equation}
\end{linenomath*}

Putting a GP prior on the vertical Pu mass profile, $\textbf{p}$, helps to enforce its smoothness. Here for computational tractability we assign a GP prior on $\textbf{y} = \log \left(\textbf{p}\right)$. Assuming a constant mean for $y$, $\mu_{\rm GP}$, we thus have

\begin{linenomath*}
	\begin{equation}
	y \sim \mathcal{GP} \left(\mu_{\rm GP},k\left(y , y'\right)\right)
	\label{mcmc10}
	\end{equation}
\end{linenomath*}

There are many possibilities for the choice of the covariance kernel $k\left(\cdot,\cdot\right)$. In this study we chose the squared exponential or radial basis function (RBF) kernel which is the de-facto default kernel for GPs. This kernel is infinitely differentiable, which means that the GP with this covariance function is very smooth \citep[][]{Rasmussen-Williams2006}

\begin{linenomath*}
	\begin{equation}
	k_{\rm RBF}\left(y, y'\right) = \sigma^2_{\rm GP} \rm{exp} \left[\displaystyle-\frac{\left(\it y - y'\right)}{2\tau^2_{\rm GP}}\right],\\
	\label{mcmc11}
	\end{equation}
\end{linenomath*}
where $\sigma^2_{\rm GP}$ and $\tau_{\rm GP}$ are the so-called kernel variance and lengthscale of the GP. Jointly inferring $\mu_{\rm GP}$, $\sigma^2_{\rm GP}$ and $\tau_{\rm GP}$ along with $\textbf{y}$ and the other unknown variables will induce a degree of smoothness in the $\textbf{y}$ profile that is consistent with the measurement data $\textbf{d}$. In other words, the MCMC sampling will pick up $\textbf{y}$ profiles that are as smooth as possible while still leading to an appropriate fit of $F\left(\bm{\uptheta}\right)$ to $\textbf{d}$.

Under the GP prior model, the prior distribution of $\textbf{y}$,  $p\left(\textbf{y}\right)$, becomes

\begin{linenomath*}
	\begin{equation}
	p\left(\textbf{y}\right) = \int \int \int p\left(\textbf{y}\right | \mu_{\rm GP}, \sigma^2_{\rm GP},\tau_{\rm GP})  p\left(\mu_{\rm GP}\right) p\left(\sigma^2_{\rm GP}\right) p\left(\tau_{\rm GP}\right)d\mu_{\rm GP}d\sigma^2_{\rm GP}d\tau_{\rm GP}.
	\label{mcmc12}
	\end{equation}
\end{linenomath*}

The $p\left(\textbf{y}\right | \mu_{\rm GP}, \sigma^2_{\rm GP},\tau_{\rm GP})$ density is known as the marginal likelihood of the GP model \citep[see][for details]{Rasmussen-Williams2006} and can be computed analytically. For our considered application we have
\begin{linenomath*}
	\begin{equation}
	\log p\left(\textbf{y}\right | \mu_{\rm GP}, \sigma^2_{\rm GP},\tau_{\rm GP}) = -\frac{1}{2}(\textbf{y} - \mu_{\rm GP})^T \textbf{K}^{-1}(\textbf{y} - \mu_{\rm GP}) - \frac{1}{2}\log|\textbf{K}|-\frac{n_y}{2}\log2\pi,
	\label{mcmc13}
	\end{equation}
\end{linenomath*}
where $\textbf{K}$ is the $n_y \times n_y$ matrix of two-point covariances given by equation (\ref{mcmc11}). With respect to the kernel hyperparameters, $\mu_{\rm GP}$, $\sigma^2_{\rm GP}$, and $\tau_{\rm GP}$, we select the following priors. 

\begin{itemize}
	\item $\log \left(\mu_{\rm GP}\right) \sim N\left(0,1\right)$. This means that for each of the 20 drum segments, the mean (total) Pu mass is assigned a lognormal prior distribution with mode, median and standard deviation of $\exp \left( -1\right)$ g, 1 g and 2.16 g, respectively. 
	\item $\log \left(\sigma^2_{\rm GP}\right) \sim N\left(0,1\right)$, which is deemed wide enough to allow for the sampled $\textbf{y}$ values to depart from $\mu_{\rm GP}$ by orders of magnitude if necessary for fitting the measurement data $\textbf{d}$.
	\item $p\left(\tau_{\rm GP}\right) \sim U\left(0.5,20\right)$. The $\tau_{\rm GP}$ parameter is the lengthscale of the GP prior, also loosely called correlation length, and controls how much two values, $y$ and $y'$ separated by a given distance will be correlated. Here the considered distance measure is the number of segments and by using a $U\left(0.5,20\right)$ prior distribution we thus allow $\tau_{\rm GP}$ to vary from half a segment to the entire profile length of 20 segments. For the considered Gaussian kernel in equation (\ref{mcmc11}), $\tau_{\rm GP}$ = 0.5 means a negligibly small spatial correlation of about 0.14 between adjacent segments while $\tau_{\rm GP}$ = 20 translates into a spatial correlation of about 1 between adjacent segments and about 0.61 between the top (segment 20) and bottom (segment 1) of the drum.
\end{itemize}

\subsubsection{Prior for the $^{\rm 244}$Cm profile}
\label{cm244_prior}

\begin{figure}[hbt!]
	\noindent\hspace{0cm}\includegraphics[width=35pc]{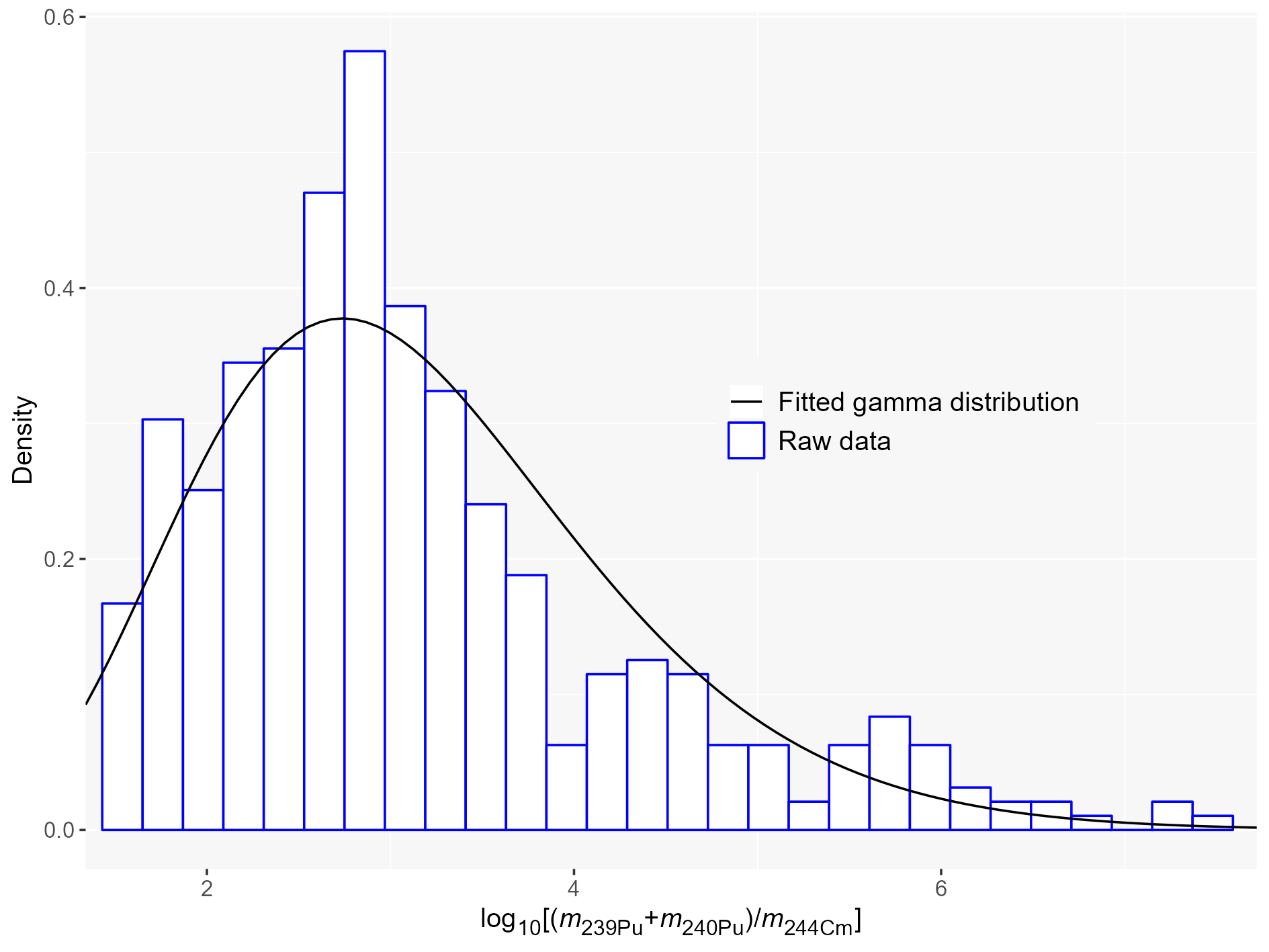}
	\caption{Distribution of the quantity $\log_{10} \left[\left(m_{\rm 239Pu}+m_{\rm 240Pu}\right)/m_{\rm 244Cm}\right]$ from the  NEA-SFCOMPO dataset and corresponding fitted gamma distribution.} 
	\label{fig2}
\end{figure}

For inference of the 20 $^{\rm 244}$Cm masses, we devise a prior based on the NEA-SFCOMPO 2.0 dataset \citep[available online, see][]{Michel-Sendis2017}. The NEA-SFCOMPO database contains $^{\rm 239}$Pu, $^{\rm 240}$Pu and $^{\rm 244}$Cm experimentally measured masses from fuel samples irradiated in power reactors over the past 50 years. It basically includes all reactor and fuel types with very low to high burnup. From this databaset, it is observed that the quantity $\log_{10} \left[\left(m_{\rm 239Pu}+m_{\rm 240Pu}\right)/m_{\rm 244Cm}\right]$ approximately follows a gamma distribution with shape parameter 7.88 and rate parameter 2.51 (Figure \ref{fig2}). We therefore infer the $cr$ coefficient with a gamma prior, $p\left(cr\right)$ = $\Upgamma\left(7.88, 2.51\right)$ and calculate the corresponding $^{\rm 244}$Cm mass in a given segment, i, as $\displaystyle m_{\rm 244Cm, i} = \frac{\left(m_{\rm 239Pu, i} + m_{\rm 240Pu, i}\right)}{10^{cr}}$ where $m_{\rm 239Pu, i}$ and $m_{\rm 240Pu, i}$ are jointly inferred along $cr$ and the other considered variables by the MCMC sampling (see section \ref{inf_var}).

\FloatBarrier

\subsubsection{Other priors}
\label{other_priors}

Alternatively to using a GP prior for $\textbf{y} = \log \left(\textbf{p}\right)$, we also test with $p\left(\textbf{y}\right) \sim N\left(0,1\right)$ and $p\left(\textbf{y}\right) \sim U\left(\log\left(\textbf{0.01}\right),\log\left(\textbf{100}\right)\right)$, respectively. Using $p\left(\textbf{y}\right) \sim N\left(0,1\right)$ is practically the same as putting a standard normal, that is, $N\left(0,1\right)$, prior on the GP mean (as done herein) and fixing $\sigma^2_{\rm GP}$ = 1 and $\tau_{\rm GP}$ = 0 in equation (\ref{mcmc11}). Furthermore, using $p\left(\textbf{y}\right) \sim U\left(\log\left(\textbf{0.01}\right),\log\left(\textbf{100}\right)\right)$ basically signifies that the Pu mass in any given segment, $p$, is allowed to vary between 0.01 g and 100 g while lower masses are preferred over larger ones if consistent with $\textbf{d}$.

The prior distributions for the 20 $^{\rm 241}$Am to Pu mass ratios, $\textbf{a}$, 20 $^{\rm 237}$Np to Pu mass ratios, $\textbf{n}$ and 20 $^{\rm 235}$U to Pu mass ratios, $\textbf{u}$, are all taken as bounded uniform distributions, of the which the lower and upper bounds are chosen on the basis of the isotopic composition analysis applied to the ISOCS gamma measurement described in section \ref{isocs_gamma}. We use $p\left(\textbf{a}\right) = U\left(\textbf{0.01},\textbf{0.15}\right)$, $p\left( \textbf{n}\right) = U\left(\textbf{0.0001},\textbf{0.01}\right)$ and $p\left(\textbf{u}\right) = U\left(\textbf{0.01},\textbf{0.15}\right)$.

Since for technical reasons we cannot assign Poisson priors to the (SGS) background continuum counts (SGS), $\textbf{b}$, we use instead an uncorrelated and independent normal prior distribution with mean and variance vectors both equal to the measured count values: $p\left(\textbf{b}\right) = N\left(\textbf{b},\textbf{C}_b\right)$ with $\textbf{C}_b$ a diagonal matrix with the $b_1,\cdots,b_{240}$ values as diagonal elements. A $N\left(x,x\right)$ distribution provides an increasingly accurate approximation to  $Pois\left(x\right)$ as $x$ increases \citep[see, e.g.,][]{Barbour1992}. The approximation is commonly deemed excellent for $x > 1000$ and reasonably accurate for $x > 10$ \citep[][]{SOCR-UCLA}.

Lastly, the prior distribution for the 20-dimensional $\bm{\uplambda}$ vector is taken as a bounded uniform distribution: $U\left(\textbf{0},\textbf{1}\right)$.

\subsection{Sampled posterior}

We assume the prior distributions for $\textbf{V}$, $\textbf{y}$, $\textbf{a}$, $cr$, $\textbf{n}$, $\textbf{u}$,  $\bm{\uplambda}$ and $\textbf{b}$ to be independent. Using Bayesian hierarchical modeling of $p\left(\textbf{V}\right)$ and a Gaussian process prior for $\textbf{y} = \log \left(\textbf{p}\right)$, the sampled posterior parameter distribution becomes
\begin{linenomath*}
	\begin{equation}
	\begin{gathered}
	p\left(\textbf{V},\textbf{y}, \textbf{a}, cr, \textbf{n} , \textbf{u}, \bm{\uplambda}, \textbf{b}, \bm{\upalpha}, \mu_{\rm GP}, \sigma^2_{\rm GP},\tau_{\rm GP} | \textbf{d}, \textbf{E}_{\rm SGS}^h,\textbf{E}_{\rm SGS}^p, \textbf{e}_{\rm PNCC}^h,\textbf{e}_{\rm PNCC}^p\right) \propto \\
	p\left(\textbf{d}, \textbf{E}_{\rm SGS}^h,\textbf{E}_{\rm SGS}^p, \textbf{e}_{\rm PNCC}^h,\textbf{e}_{\rm PNCC}^p \right | \textbf{V},\textbf{y}, \textbf{a}, cr, \textbf{n} , \textbf{u}, \bm{\uplambda}, \textbf{b}) \times \\ p\left(\textbf{V} | \bm{\upalpha}\right)p\left(\bm{\upalpha}\right)p\left(\textbf{y}\right | \mu_{\rm GP}, \sigma^2_{\rm GP},\tau_{\rm GP})  p\left(\mu_{\rm GP}\right) p\left(\sigma^2_{\rm GP}\right) p\left(\tau_{\rm GP}\right) \times \\ p\left(\textbf{a}\right) p\left(cr\right) p\left(\textbf{n}\right) p\left(\textbf{u}\right) p\left(\bm{\uplambda}\right) p\left(\textbf{b}\right).
	\label{mcmc15}
	\end{gathered}
	\end{equation}
\end{linenomath*}

No analytical solution of the posterior distribution in equation (\ref{mcmc15}) can be derived and we therefore sample from this distribution by MCMC simulation \citep[see][]{Gelman2014} using the efficient HMC sampler \citep[see][for an extensive description of the HMC algorithm]{Neal2011,Betancourt2018}. Convergence of the MCMC to the posterior target is monitored by means of the potential scale reduction factor, $\hat{R}$ \citep{Gelman-Rubin1992,Gelman2014}, using eight interacting Markov chains evolved in parallel (see next section). The $\hat{R}$ statistic compares for each parameter of interest the average within-chain variance to the variance of all the Markov chains mixed together. The closer the values of these two variances, the closer to one the value of the $\hat{R}$ diagnostic. To declare convergence to a limiting distribution we require that the values of $\hat{R}$ are jointly smaller than 1.1 for all sampled variables.

\subsection{Software implementation}
\label{software}

Similarly as in \citet{Laloy2021}, we used the open-source greta package \citep[][]{Golding2019} to perform the HMC-based MCMC sampling. The greta package provides a probabilistic programming language embedded in R \citep{Rsoftware} interfacing to some of the MCMC sampling algorithms implemented in the Tensorflow-probability package \citep[TFP,][]{Dillon2017}, which itself relies on the Tensorflow (TF) machine learning platform \citep[][]{tensorflow2016}. The most useful MCMC sampler available through greta and used herein is HMC \citep[][]{Neal2011,Betancourt2018}. This TFP-based HMC implementation can evolve several Markov chains in parallel on both CPUs and GPUs, with the different chains exchanging information during warmup to speedup convergence. In this study, we evolved 8 interacting Markov chains in parallel over 6 CPUs. Additionally, most of the pre- and post-processing was performed with the tidyverse collection of packages \citep{tidyverse}, together with a few other specific packages \citep[mainly by][]{coda,ggdist,patchwork}.

\section{Results and discussion}
\label{results}

We first investigate the behavior of our proposed inference method on a synthetic experiment for which the ``truth" is thus known, before applying it to the same real waste drum as considered by \citet{Laloy2021}. 

\subsection{Synthetic experiment}
\label{synth_exp}

The reference or ``true" drum was built by as follows. The true total Pu mass (g) profile across the 20 segments, $\textbf{p}_{true}$, was taken as
\begin{linenomath*}
	\begin{equation}
	\begin{gathered}
	\textbf{p}_{true} = 0.15 + \left[\sin \left(0.5\textbf{x}\right)+1 \right] \times 1.175 \\
	\text{with}\ \textbf{x} = \left[1, 2, \cdots, 20\right].
	\end{gathered}
	\label{true_ptot}
	\end{equation}
\end{linenomath*}

Furthermore, the true $\lambda$ value was set to 1 (homogeneous source) in every segment except for segment 16 where it was set to 0 (point source). True values for all other inferred variables were randomly sampled from their respective prior distributions. 

The true SGS and PNCC measurements were then created by randomly perturbing the forward-simulated data corresponding to the true parameter values. More specifically, the $i=1,\cdots,240$ simulated net counts (associated with SGS), $c^{net}_{true,i}$, were used as mean parameters of 240 Poisson distributions from which the 240 observed net counts were drawn: $c^{net}_{obs,i} \propto Pois\left(c^{net}_{true,i}\right)$. The 240 observed background continuum counts, $b_{obs,i}$, were set to the measured ones for the real case (section \ref{real_exp}). In addition, the observed number of reals per second (associated with PNCC), $rs_{obs}$, was drawn from a normal distribution centered around the true simulated number of reals per second, $rs_{true}$, and relative standard deviation of 5\%. Note that due to the added errors in the observed net counts and reals per second, no perfect inverse solution exists which is required for Bayesian inference.

We first consider Bayesian hierarchical modeling of the prior uncertainty in the 20 Pu isotopic vectors (equation (\ref{mcmc7}) in section \ref{alpha_prior}) and putting a GP prior model on the total Pu mass across segments (section \ref{gp_prior}). This inversion strategy is referred to as strategy 1. The HMC was ran for 10,000 warmup iterations and 5000 consecutive sampling iterations, thereby providing 40,000 posterior samples. 

\begin{figure}[hbt!]
	\noindent\hspace{0cm}\includegraphics[width=35pc]{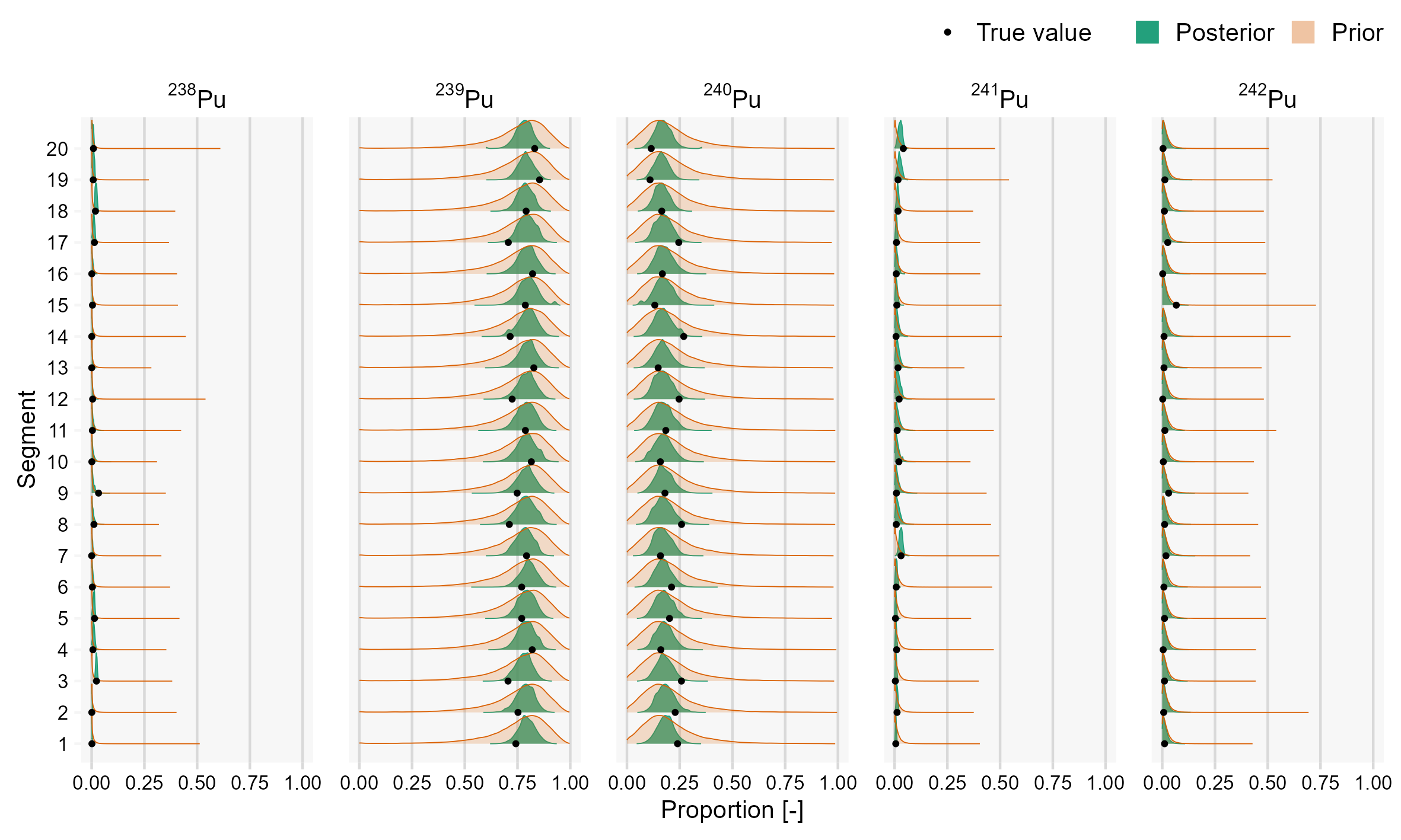}
	\caption{Prior and posterior Pu isotopic vector distributions in each of the 20 segments for the synthetic test case and strategy 1. Strategy 1 uses Bayesian hierarchical modeling of the Pu isotopic vector uncertainty and a GP prior model for the logarithm of the total Pu mass per segment. Segment numbering goes from bottom (1) to top (20).} 
	\label{fig3}
\end{figure}

\begin{figure}[hbt!]
	\noindent\hspace{-1.2cm}\includegraphics[width=45pc]{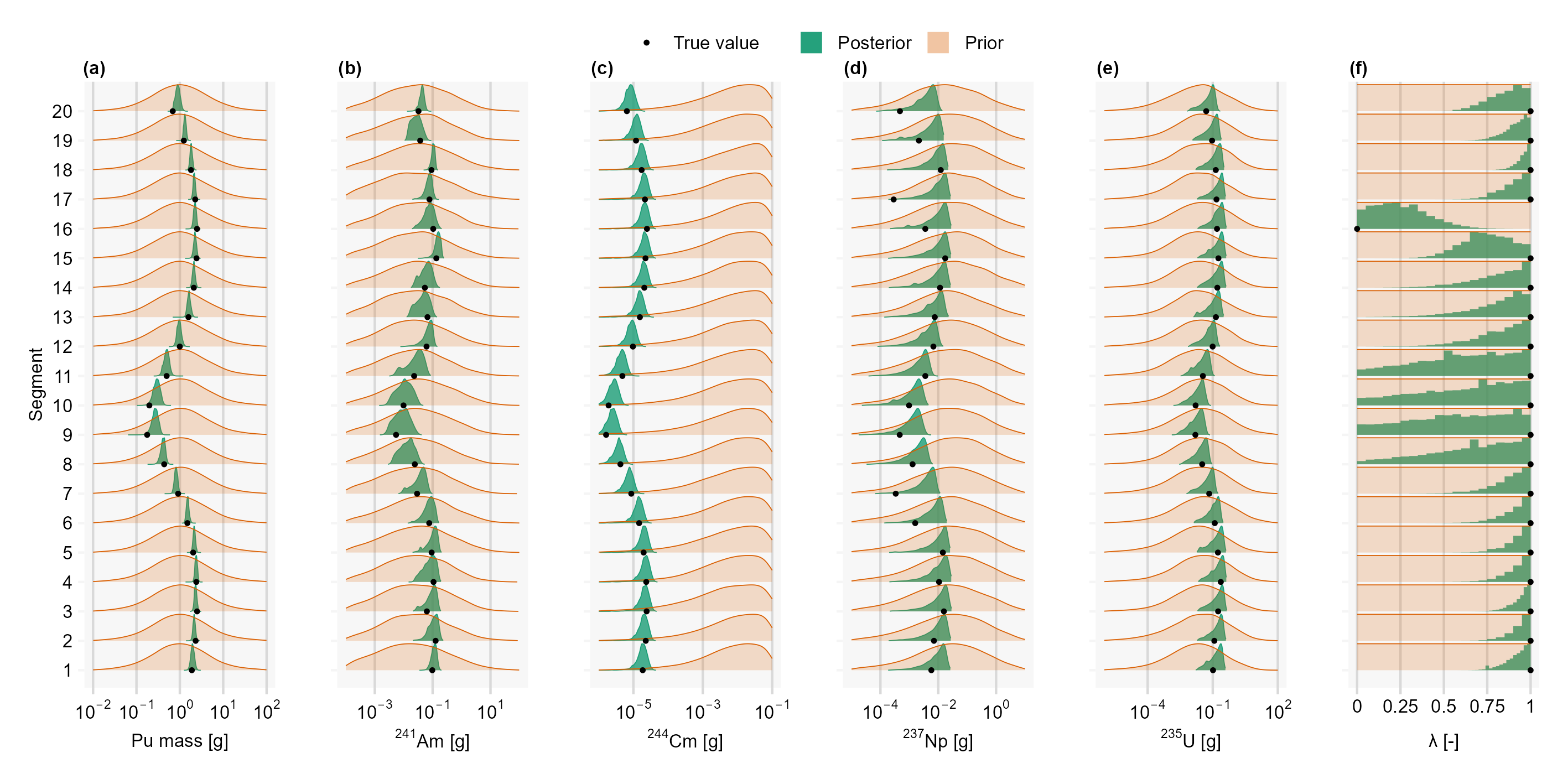}
	\caption{Prior and posterior (a) Pu mass, (b) $^{\rm 241}$Am mass, (c) $^{\rm 244}$Cm, (d) $^{\rm 237}$Np, (e) $^{\rm 235}$U and (f) $\lambda$ distributions for the synthetic test and strategy 1. Strategy 1 uses Bayesian hierarchical modeling of the Pu isotopic vector uncertainty and a GP prior model for the logarithm of the total Pu mass per segment. The black dots denote the true values used to generate the SGS (gross count) and PNCC (reals per second) data. The outcome of equation (\ref{true_ptot}) is the distribution of black dots in subplot (a). Note that the $x$-axes of subplots (a - e) have a base 10 logarithmic scale. When plotted with this scale, all prior distributions in subplots (a - e) are symmetric but, for visual convenience, the corresponding $x$-axes do not cover the full prior range (especially for subplots b, c and d). For the $\lambda$ variable, we show a posterior histogram instead of a kernel density estimate. This is because when applied to a bounded data sample, kernel density smoothing tends to create artifacts near the bounds. Segment numbering goes from bottom (1) to top (20).} 
	\label{fig4}
\end{figure}

Figures \ref{fig3} and \ref{fig4} depict the posterior distributions of the Pu isotopic vectors and total nuclide masses, respectively, across the considered 20 segments. It is seen that the posterior distribution is overall well resolved and always contains the true parameter value. With respect to the posterior $\bm{\uplambda}$ distribution (source distribution uncertainty in each segment), we find that the posterior uncertainty is larger in case of lower total radioactive content in the segment (Figure \ref{fig4}f). This is likely due to the fact that a lower amount of Pu produces less net counts (for gamma-emitting isotopes) and reals per seconds (for the odd Pu isotopes) which reduces the available information about $\bm{\uplambda}$.

\begin{figure}[hbt!]
	\noindent\hspace{0cm}\includegraphics[width=35pc]{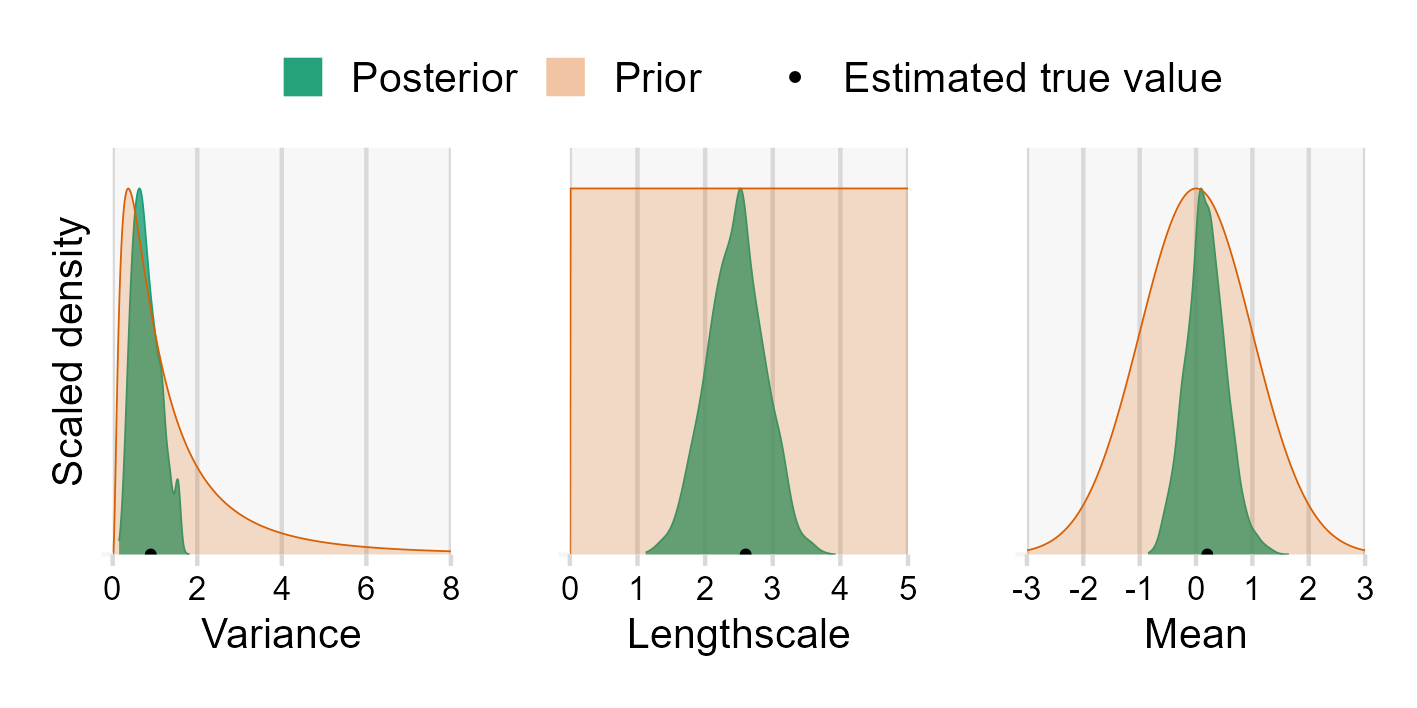}
	\caption{Prior and posterior (a) $\sigma^2_{\rm GP}$, (b) $\tau_{\rm GP}$ and (c) $\mu_{\rm GP}$ distributions for the synthetic test case and strategy 1. Strategy 1 uses Bayesian hierarchical modeling of the Pu isotopic vector uncertainty and a GP prior model for the logarithm of the total Pu mass per segment.} 
	\label{fig5}
\end{figure}

Figure \ref{fig5} shows the posterior distribution of the inferred GP parameters, $\mu_{\rm GP}$, $\sigma^2_{\rm GP}$ and $\tau_{\rm GP}$, together with the associated ``estimated" true values obtained by fitting our GP model to $\log \left(\textbf{p}_{true}\right)$ directly. The marginal posterior GP parameter distribution appears to be in good agreement with the estimated truth.

\FloatBarrier

\begin{figure}[hbt!]
	\noindent\hspace{-1.2cm}\includegraphics[width=45pc]{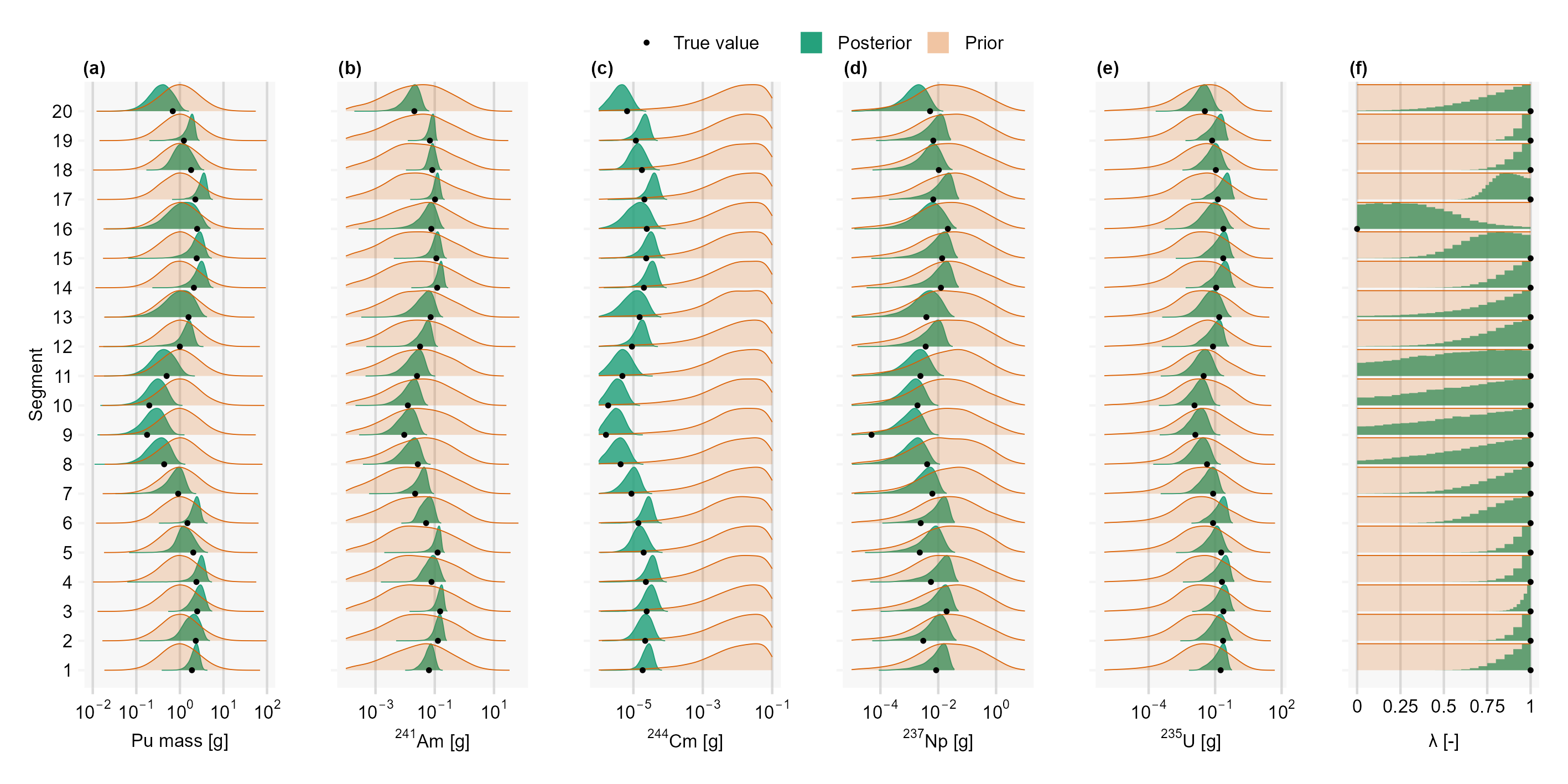}
	\caption{Prior and posterior (a) Pu mass, (b) $^{\rm 241}$Am mass, (c) $^{\rm 244}$Cm, (d) $^{\rm 237}$Np, (e) $^{\rm 235}$U and (f) $\lambda$ distributions for the synthetic test case and strategy 2. Strategy 2 relies on Bayesian hierarchical modeling of the Pu isotopic vector uncertainty and puts a standard normal prior on the logarithm of the total Pu mass per segment. The black dots denote the true values used to generate the SGS (gross count) and PNCC (reals per second) data.  The outcome of equation (\ref{true_ptot}) is the distribution of black dots in subplot (a). Note that the $x$-axes of subplots (a - e) have a base 10 logarithmic scale. When plotted with this scale, all prior distributions in subplots (a - e) are symmetric but, for visual convenience, the corresponding $x$-axes do not cover the full prior range. For the $\lambda$ variable, we show a posterior histogram instead of a kernel density estimate. This is because when applied to a bounded data sample, kernel density smoothing tends to create artifacts near the bounds. Segment numbering goes from bottom (1) to top (20).} 
	\label{fig6}
\end{figure}

\begin{figure}[hbt!]
	\noindent\hspace{-1.2cm}\includegraphics[width=45pc]{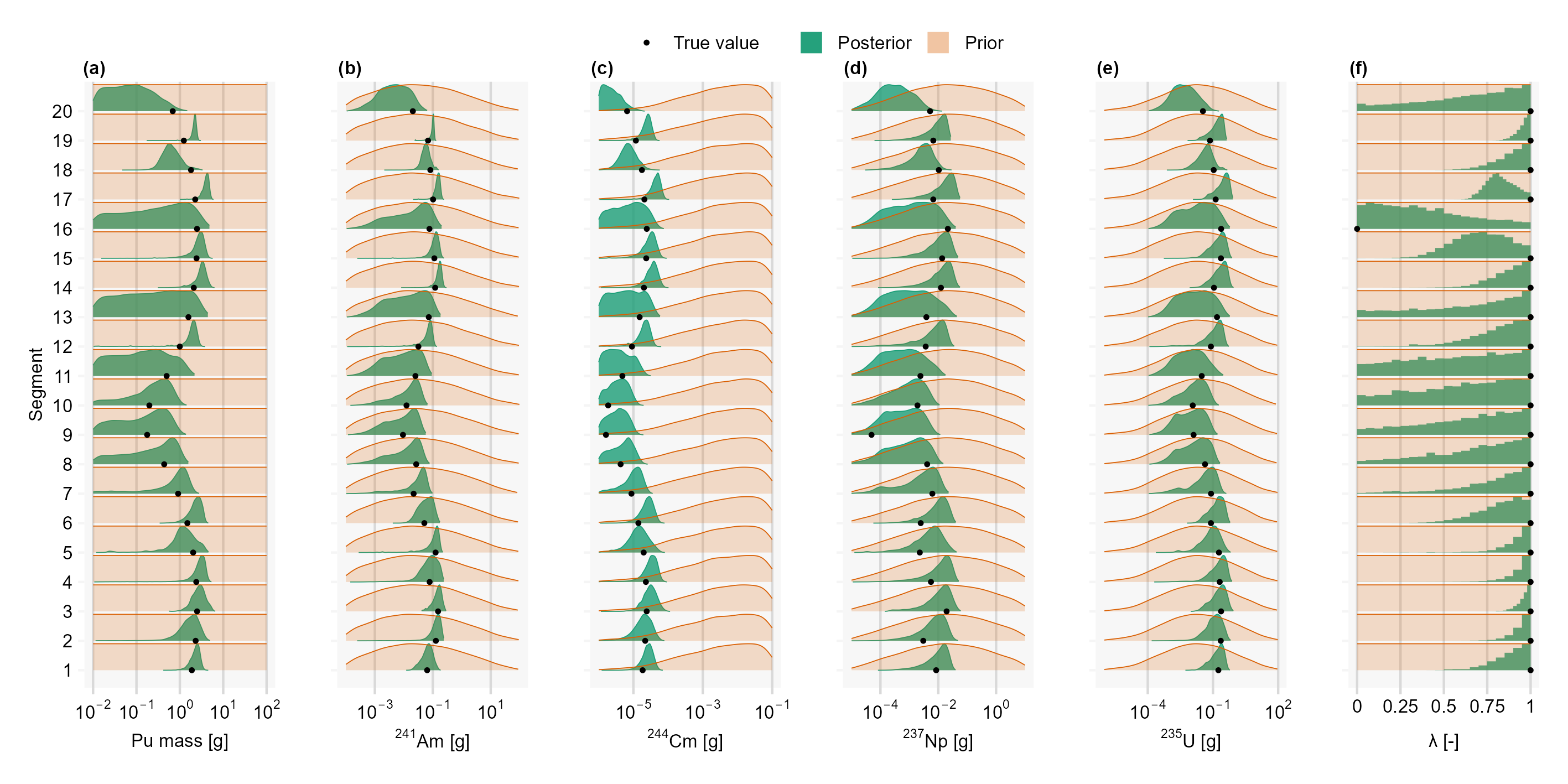}
	\caption{Prior and posterior (a) Pu mass, (b) $^{\rm 241}$Am mass, (c) $^{\rm 244}$Cm, (d) $^{\rm 237}$Np, (e) $^{\rm 235}$U and (f) $\lambda$ distributions for the synthetic test case and strategy 3. Strategy 3 relies on Bayesian hierarchical modeling of the Pu isotopic vector uncertainty and puts an uniform prior on the logarithm of the total Pu mass per segment. The black dots denote the true values used to generate the SGS (gross count) and PNCC (reals per second) data.  The outcome of equation (\ref{true_ptot}) is the distribution of black dots in subplot (a). Note that the $x$-axes of subplots (a - e) have a base 10 logarithmic scale. When plotted with this scale, all prior distributions in subplots (a - e) are symmetric but, for visual convenience, the corresponding $x$-axes do not cover the full prior range. For the $\lambda$ variable, we show a posterior histogram instead of a kernel density estimate. This is because when applied to a bounded data sample, kernel density smoothing tends to create artifacts near the bounds. Segment numbering goes from bottom (1) to top (20).} 
	\label{fig7}
\end{figure}

We now turn attention to the case where no GP prior model is used for the total Pu mass but either a standard normal prior or a loguniform prior bounded between 0.1 g and 100 g in each segment. We refer to these two inversion strategies as strategies 2 and 3, respectively. As written above, using a standard normal prior (strategy 2) is equivalent to using the GP prior model described in equations (\ref{mcmc10}) - (\ref{mcmc11}) (strategy 1) but with fixing $\sigma^2_{\rm GP}$ = 1 and $\tau_{\rm GP}$ = 0. In case of the loguniform prior (strategy 3), the warmup period needed to be extended to 20,000 iterations to achieve $\hat{R}$-convergence. Figures \ref{fig6} presents the posterior mass and $\bm{\uplambda}$ distributions in each segment corresponding to strategy 2 while Figure \ref{fig7} displays the same distributions for strategy 3. These distributions are fairly (strategy 2) to strongly (strategy 3) wider than when a GP prior model is used for the total Pu mass (strategy 1, Figure \ref{fig4}). Furthermore, although narrower the posterior distributions associated with strategy 1 always include the true values. This nicely illustrates the benefits of GP-based regularization to reduce posterior uncertainty. 

\FloatBarrier

\begin{figure}[hbt!]
	\noindent\hspace{-1.2cm}\includegraphics[width=45pc]{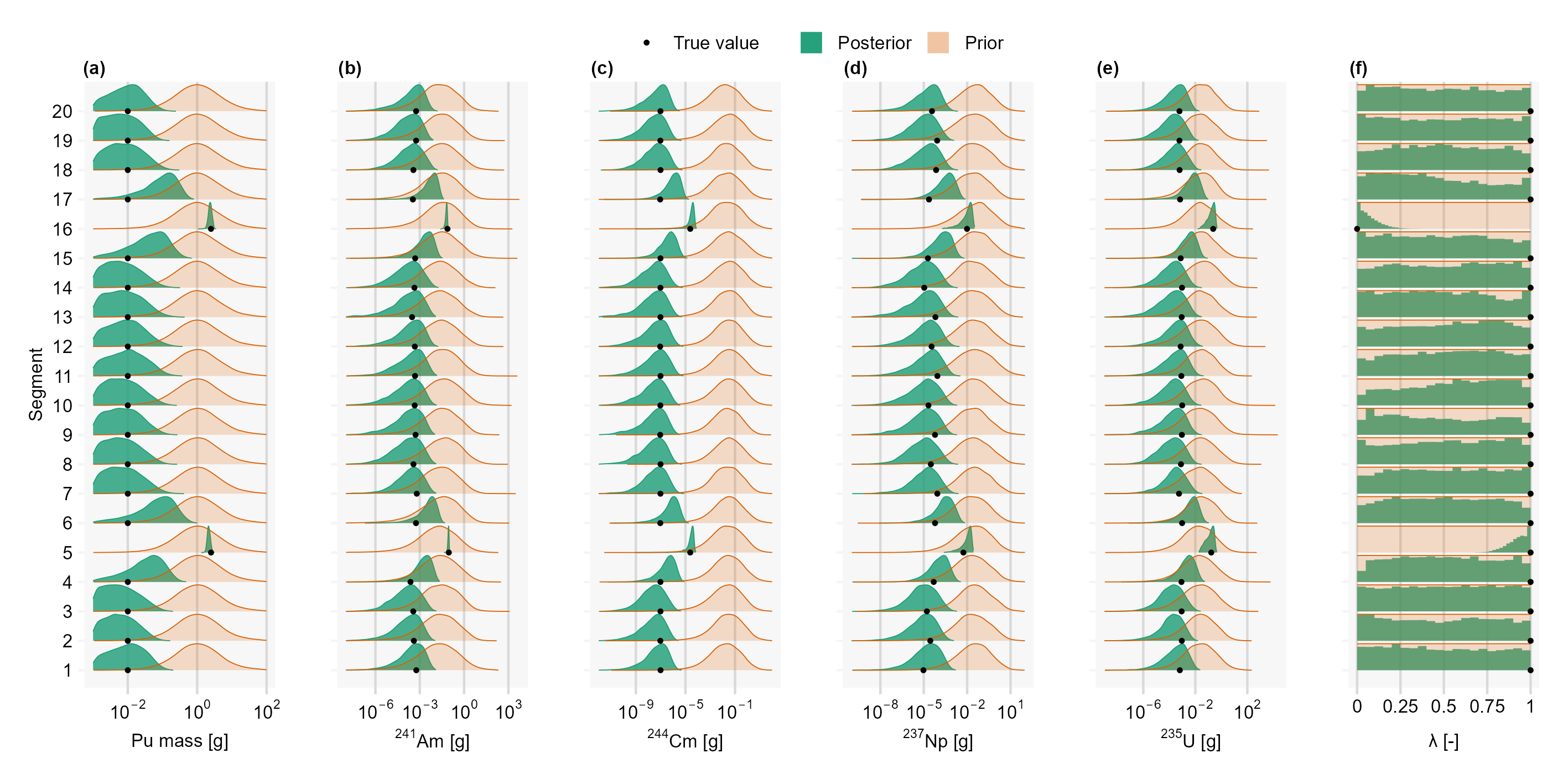}
	\caption{Prior and posterior (a) Pu mass, (b) $^{\rm 241}$Am mass, (c) $^{\rm 244}$Cm, (d) $^{\rm 237}$Np, (e) $^{\rm 235}$U and (f) $\lambda$ distributions for the synthetic test case and strategy 1, when using a ``flat and spiky" true Pu mass profile. The black dots denote the true values used to generate the SGS (gross count) and PNCC (reals per second) data. Note that the $x$-axes of subplots (a - e) have a base 10 logarithmic scale. When plotted with this scale, all prior distributions in subplots (a - e) are symmetric but, for visual convenience, the corresponding $x$-axes do not cover the full prior range. For the $\lambda$ variable, we show a posterior histogram instead of a kernel density estimate. This is because when applied to a bounded data sample, kernel density smoothing tends to create artifacts near the bounds. Segment numbering goes from bottom (1) to top (20).} 
	\label{fig8}
\end{figure}

To verify whether using a GP prior model remains useful when $\log \left(\textbf{p}_{true}\right)$ does not at all obey a GP, we repeated the exercise with a GP prior put on $\log \left(\textbf{p}\right)$ (strategy 1) but for a case where $p_{true}$ is set to 0.015 g in every segment but segments 5 and 16 where it is set to 2.5 g. Hence, the $\log \left(\textbf{p}_{true}\right)$ profile is now flat with two spikes (in segments 5 and 16). Such a ``flat and spiky" profile is arguably far from Gaussian. Figure \ref{fig8} presents the corresponding posterior mass and $\bm{\uplambda}$ distributions in each segment. It is observed that the inferred posterior mass profiles always include the true mass values, though at the sharp interfaces (upper and lower neighboring segments of segment 5 and 16), most of the probability mass is put at somewhat too large values (see posterior mass distributions in segments 4 and 6 and segments 15 and 17, respectively). This can be fully mitigated by using a log-uniform rather uniform prior for $\tau_{\rm GP}$, together with a lower bound of 0.01 instead of 0.5 (not shown). Nevertheless, such setting of the prior $\tau_{\rm GP}$ distribution assumes that some prior knowledge of the true $\log \left(\textbf{p}_{true}\right)$ profile is available. Overall, we conclude from this last synthetic test that even if the true Pu mass profile does not at all show smooth transitions between segments, using a GP prior still allows for finding a regularized solution that is reasonably accurate. 

\FloatBarrier

\subsection{Real drum}
\label{real_exp}

For the real data inversion, our reference strategy is strategy 1 which therefore includes hierarchical modeling of the Pu isotopic vector uncertainty and putting a GP prior model on the total Pu mass. 

\begin{figure}[hbt!]
	\noindent\hspace{0cm}\includegraphics[width=35pc]{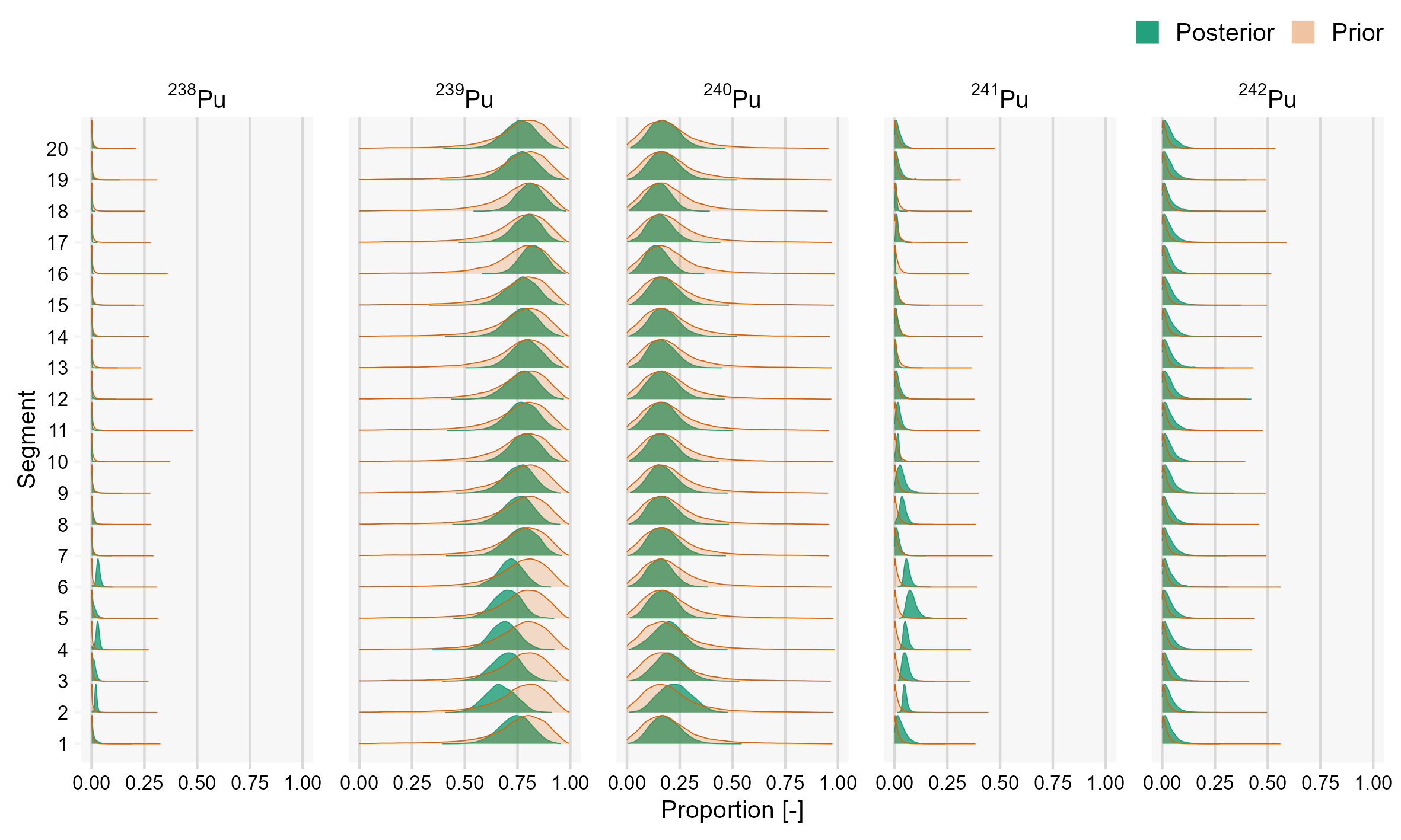}
	\caption{Prior and posterior Pu isotopic vector distributions in each of the 20 segments for the real test case and strategy 1. Strategy 1 uses Bayesian hierarchical modeling of the Pu isotopic vector uncertainty and a GP prior model for the logarithm of the total Pu mass per segment. Segment numbering goes from bottom (1) to top (20).} 
	\label{fig9}
\end{figure}

\begin{figure}[hbt!]
	\noindent\hspace{-1.2cm}\includegraphics[width=45pc]{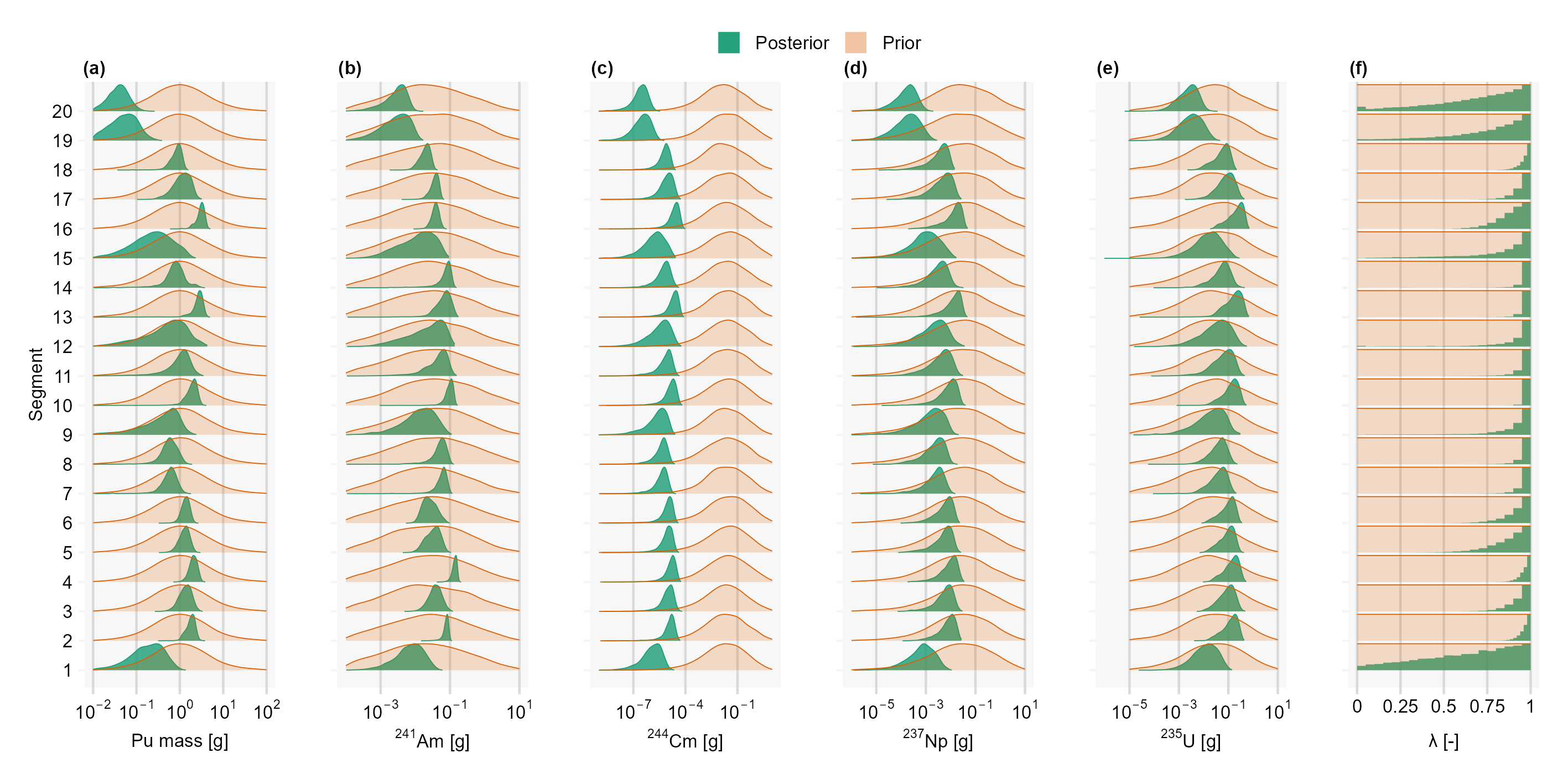}
	\caption{Prior and posterior (a) Pu mass, (b) $^{\rm 241}$Am mass, (c) $^{\rm 244}$Cm, (d) $^{\rm 237}$Np, (e) $^{\rm 235}$U and (f) $\lambda$ distributions for the real test case and strategy 1. Strategy 1 uses Bayesian hierarchical modeling of the Pu isotopic vector uncertainty and a GP prior model for the logarithm of the total Pu mass per segment. Note that the $x$-axes of subplots (a - e) have a base 10 logarithmic scale. When plotted with this scale, all prior distributions in subplots (a - e) are symmetric but, for visual convenience, the corresponding $x$-axes do not cover the full prior range. For the $\lambda$ variable, we show a posterior histogram instead of a kernel density estimate. This is because when applied to a bounded data sample, kernel density smoothing tends to create artifacts near the bounds. Segment numbering goes from bottom (1) to top (20).} 
	\label{fig10}
\end{figure}

\begin{figure}[hbt!]
	\noindent\hspace{0cm}\includegraphics[width=35pc]{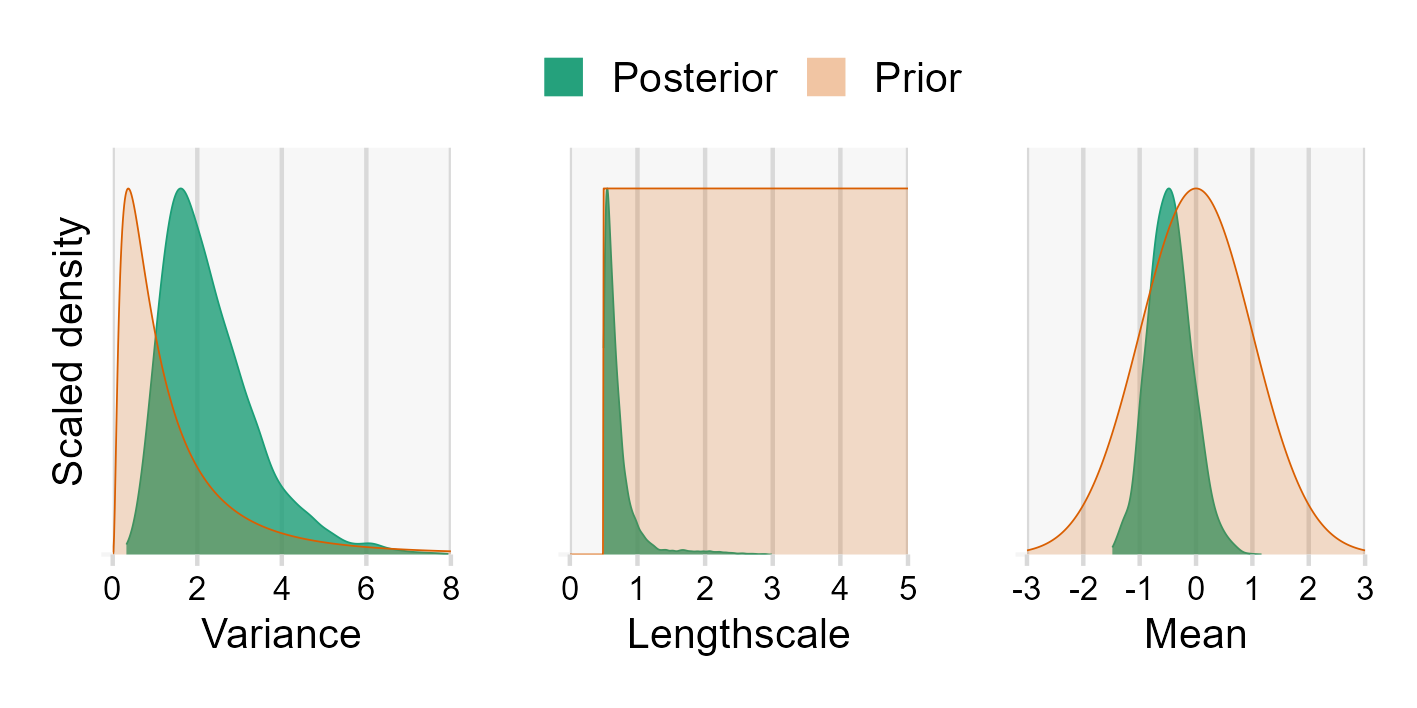}
	\caption{Prior and posterior (a) $\sigma^2_{\rm GP}$, (b) $\tau_{\rm GP}$ and (c) $\mu_{\rm GP}$ distributions for the real test case and strategy 1. Strategy 1 uses Bayesian hierarchical modeling of the Pu isotopic vector uncertainty and a GP prior model for the logarithm of the total Pu mass per segment.} 
	\label{fig11}
\end{figure}

Some of the associated inversion results are illustrated in Figures \ref{fig9} (posterior distribution of the isotopic vectors), \ref{fig10} (posterior mass and $\bm{\uplambda}$ distributions) and \ref{fig11} (posterior GP parameter distributions). All sampled variables appear to be relatively well constrained, with marginal posterior distributions that are substantially narrower than their associated prior distributions. Consistently with our previous study \citep[][]{Laloy2021}, the total Pu mass is lower at the top and, to a lesser extent, at the bottom of the drum (Figure \ref{fig10}a), while the assumption of an homogeneous source distribution is strongly favored for segments 2 - 18 and more uncertain at the top (segments 19 - 20) and bottom (segment 1). As detailed in \citet[][]{Laloy2021}, the smaller posterior Pu masses in the top and bottom segments is likely caused by an incomplete drum's filling (top segments) and the fact that the bottom plate of the drum does not touch the floor, only the drum's outer bottom ring does (bottom segment).

As of the GP model parameters, the posterior lengthscale distribution is concentrated around its lower bound of 0.5. This basically means that no spatial correlation is inferred from the measurement data. This does not prevent the derived $\log \left(\textbf{p}\right)$ profile to be somewhat smooth and well resolved (Figure \ref{fig10}a). In addition, for this application we find that strategy 2 results into a posterior distribution (not shown) that resembles closely the posterior distribution achieved by strategy 1. This is mainly due to the absence of detectable spatial correlation which causes the prior distribution for $\textbf{y}$ to become relatively similar between strategies 1 and 2. As detailed earlier, for sufficiently small $\tau_{\rm GP}$ values (say $\leq$ 0.5 - 0.6), the only practical difference between the two priors is that for strategy 1 the $\sigma^2_{\rm GP}$ parameter is jointly inferred with the other variables with a lognormal prior that has a median of 1 while for strategy 2 the $\sigma^2_{\rm GP}$ parameter is fixed to 1.

\begin{figure}[hbt!]
	\noindent\hspace{-1.2cm}\includegraphics[width=45pc]{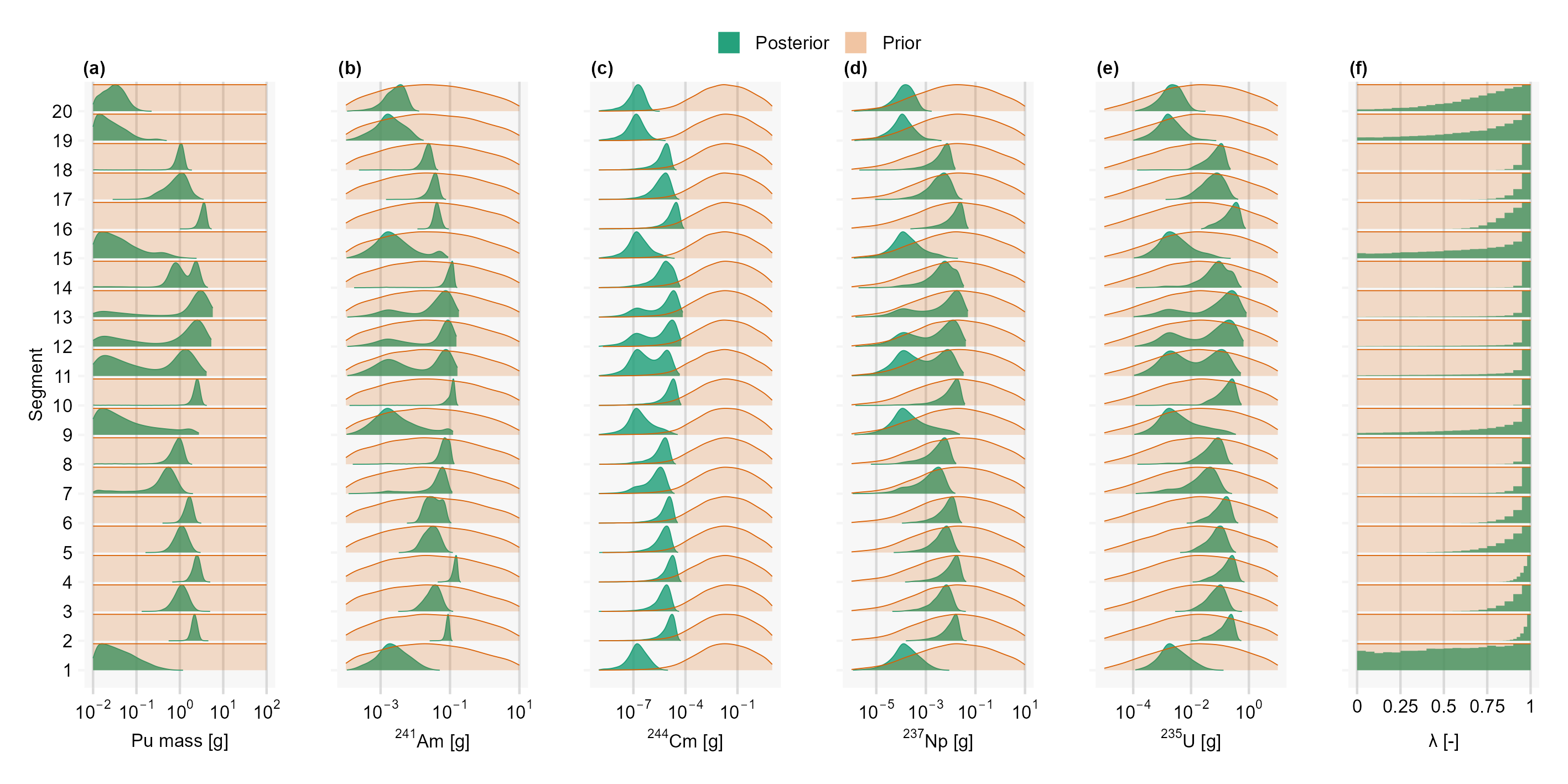}
	\caption{Prior and posterior (a) Pu mass, (b) $^{\rm 241}$Am mass, (c) $^{\rm 244}$Cm, (d) $^{\rm 237}$Np, (e) $^{\rm 235}$U and (f) $\lambda$ distributions for the real test case and strategy 3. Strategy 3 relies on Bayesian hierarchical modeling of the Pu isotopic vector uncertainty and puts an uniform prior on the logarithm of the total Pu mass per segment. Note that the $x$-axes of subplots (a - e) have a base 10 logarithmic scale. When plotted with this scale, all prior distributions in subplots (a - e) are symmetric but, for visual convenience, the corresponding $x$-axes do not cover the full prior range. For the $\lambda$ variable, we show a posterior histogram instead of a kernel density estimate. This is because when applied to a bounded data sample, kernel density smoothing tends to create artifacts near the bounds. Segment numbering goes from bottom (1) to top (20).} 
	\label{fig12}
\end{figure}

In contrast to our strategies 1 and 2, Figure \ref{fig12} illustrates that if an uniform prior is put on $\log \left(\textbf{p}\right)$ instead of a GP or standard normal prior, which as written already corresponds to inversion strategy 3, then posterior uncertainty becomes substantially larger while bimodality appears in the posterior mass distributions in some segments (see segments 11, 12 and 15). This bimodality is due to the large negative correlations between the posterior masses that appear for some adjacent segments. In addition, likewise for the synthetic case study strategy 2 incurs a larger computational cost to reach $\hat{R}$-convergence of the MCMC: 20,000 warmup iterations instead of 10,000 warmup iterations for strategies 1 and 2.

\FloatBarrier

\begin{figure}[hbt!]
	\noindent\hspace{0cm}\includegraphics[width=35pc]{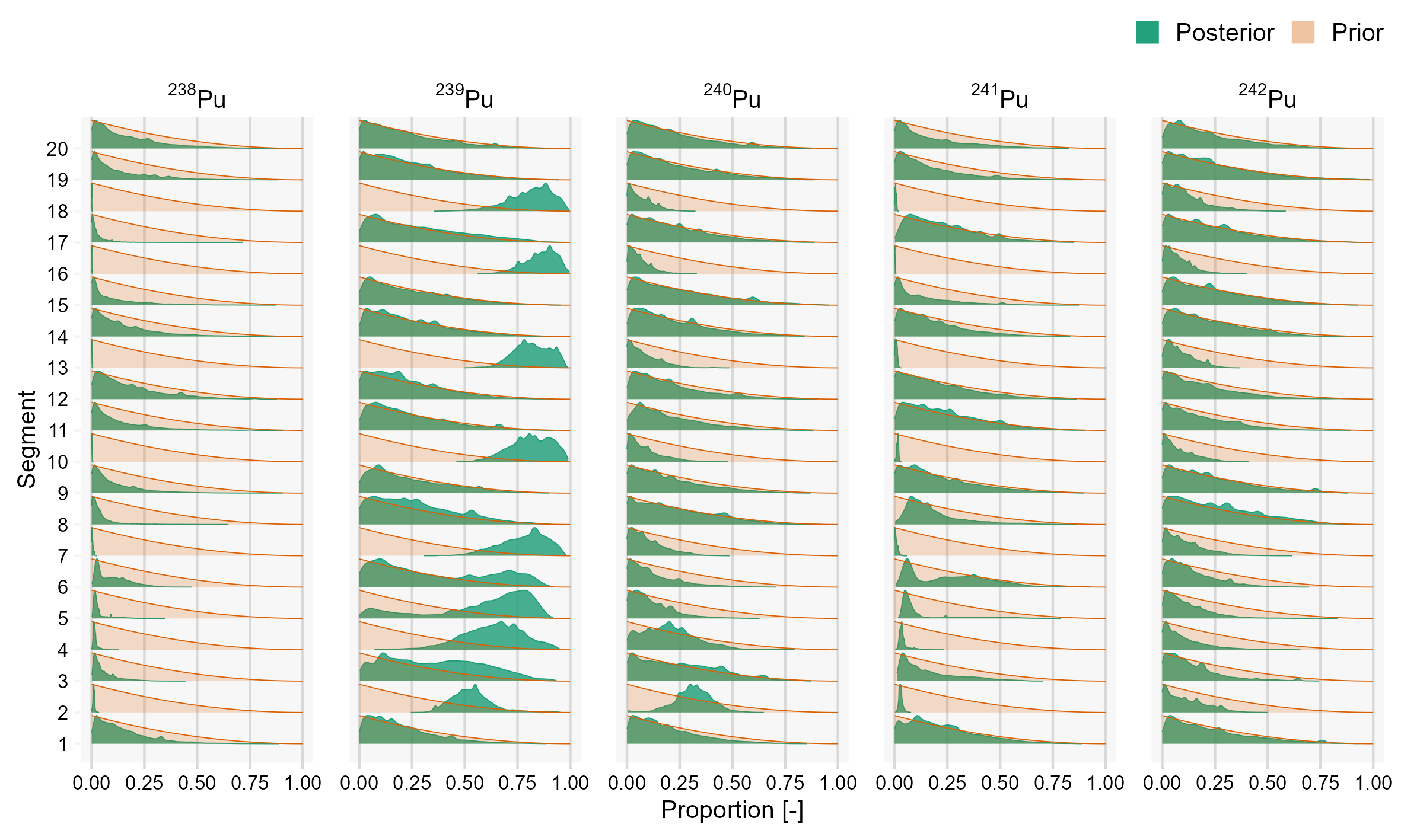}
	\caption{Prior and posterior Pu isotopic vector distributions in each of the 20 segments for the real test case and strategy 4. Strategy 4 includes putting flat Dirichlet priors on the 20 Pu isotopic vectors and putting a GP prior model on the logarithm of the total Pu mass per segment. Segment numbering goes from bottom (1) to top (20).} 
	\label{fig13}
\end{figure}

\begin{figure}[hbt!]
	\noindent\hspace{-1.2cm}\includegraphics[width=45pc]{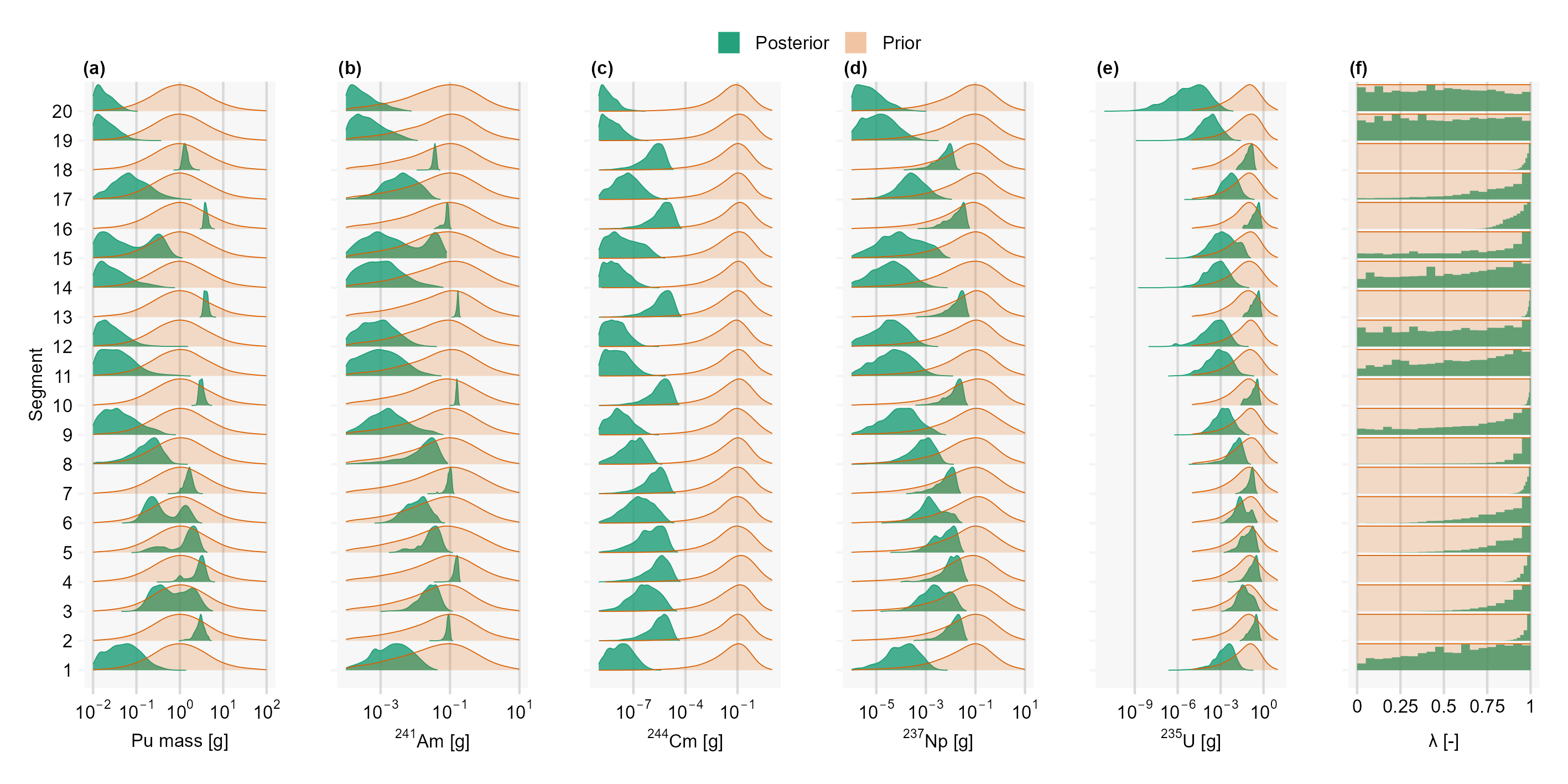}
	\caption{Prior and posterior (a) Pu mass, (b) $^{\rm 241}$Am mass, (c) $^{\rm 244}$Cm, (d) $^{\rm 237}$Np, (e) $^{\rm 235}$U and (f) $\lambda$ distributions for the real test case and strategy 4.  Strategy 4 includes putting flat Dirichlet priors on the 20 Pu isotopic vectors and putting a GP prior model on the logarithm of the total Pu mass per segment. Note that the $x$-axes of subplots (a - e) have a base 10 logarithmic scale. When plotted with this scale, all prior distributions in subplots (a - e) are symmetric but, for visual convenience, the corresponding $x$-axes do not cover the full prior range. For the $\lambda$ variable, we show a posterior histogram instead of a kernel density estimate. This is because when applied to a bounded data sample, kernel density smoothing tends to create artifacts near the bounds. Segment numbering goes from bottom (1) to top (20).} 
	\label{fig14}
\end{figure}

Figures \ref{fig13} and \ref{fig14} present the posterior Pu isotopic vectors' distribution and posterior nuclide masses' distribution, respectively, obtained when there is no prior information available on the 20 isotopic vectors and a flat Dirichlet prior is thus put on the $i = 1, \cdots, 20$ $\textbf{v}_i$ vectors, while a GP prior model is used for $\log \left(\textbf{p}\right)$. This variant is referred to as strategy 4. Herein too, the posterior mass uncertainty increases a lot for each nuclide compared to strategy 1 (Figure \ref{fig10}) and some bimodality is observed (see posterior masses in segments 5 and 15 of Figure \ref{fig14}). Furthermore, the computational effort required to achieve $\hat{R}$-convergence now becomes even larger than for the other two considered inference strategies as after 30,000 warmup iterations (followed by 5,000 sampling iterations) only 27 (12) out of the 444 sampled parameters associated with strategy 3 show a $\hat{R} < 1.1$ (1.2). The results depicted in Figures \ref{fig13} and \ref{fig14} are therefore not officially converged and should be considered with some caution, although we do not expect the main trends seen in Figures \ref{fig13} - \ref{fig14} to change drastically with more warmup iterations, only the currently observed small bumps should be smoothed out.

\FloatBarrier

\begin{figure}[hbt!]
	\noindent\hspace{-1.0cm}\includegraphics[width=40pc]{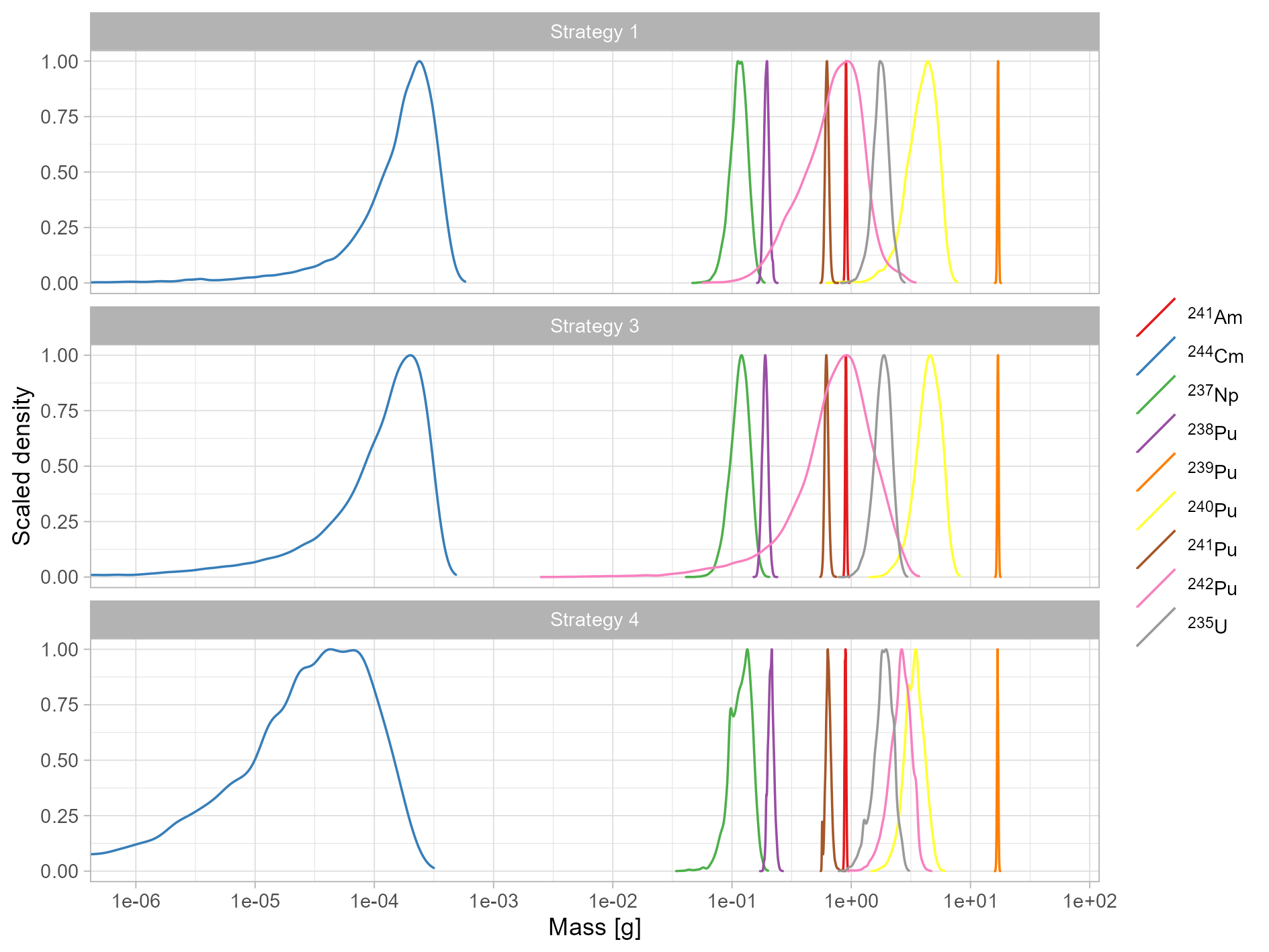}
	\caption{Posterior mass distributions over the whole drum for the considered radionuclides and MCMC inversion strategies 1, 3 and 4. For this case study, strategy 2 leads to quite similar results as strategy 1 (see main text for details).} 
	\label{fig15}
\end{figure}

Figure \ref{fig15} displays the posterior mass distributions over the drum for all considered nuclides and strategies 1, 3 and 4. We find that although strategies 3 and 4 lead to a much larger uncertainty a the segment level compared to strategy 1, strategies 3 and 4 provide similarly well resolved posterior mass distributions for the whole drum. With respect to strategy 3, the posterior mass distributions are basically the same as for strategy 1. A similar situation is observed for strategy 4 for all nuclides but $^{\rm 244}$Cm (dark blue curve) $^{\rm 240}$Pu (yellow curve) and $^{\rm 242}$Pu (pink curve). The relatively big shift towards higher $^{\rm 242}$Pu masses for strategy 4 is primarily caused by the use of a flat Dirichlet prior for the Pu isotopic vector together with the fact that $^{\rm 242}$Pu is not sensed by the (3AX-SGS) gamma measurement. The small shift toward smaller $^{\rm 240}$Pu posterior values observed for strategy 4 can also be explained by the use of a flat Dirichlet prior for the Pu isotopic vector, while an informative Dirichlet prior centered near a $^{\rm 240}$Pu mass fraction of 0.18 is used with strategies 1 and 3. With respect to $^{\rm 244}$Cm, the difference is likely caused by the difference in inferred $^{\rm 240}$Pu content as our inferred $^{\rm 244}$Cm content is co-derived from the inferred $^{\rm 239}$Pu and $^{\rm 240}$Pu masses on the one hand, and the inferred $cr$ parameter on the other hand (see section \ref{cm244_prior} for details).

\FloatBarrier

\begin{figure}[hbt!]
	\noindent\hspace{-1.2cm}\includegraphics[width=45pc]{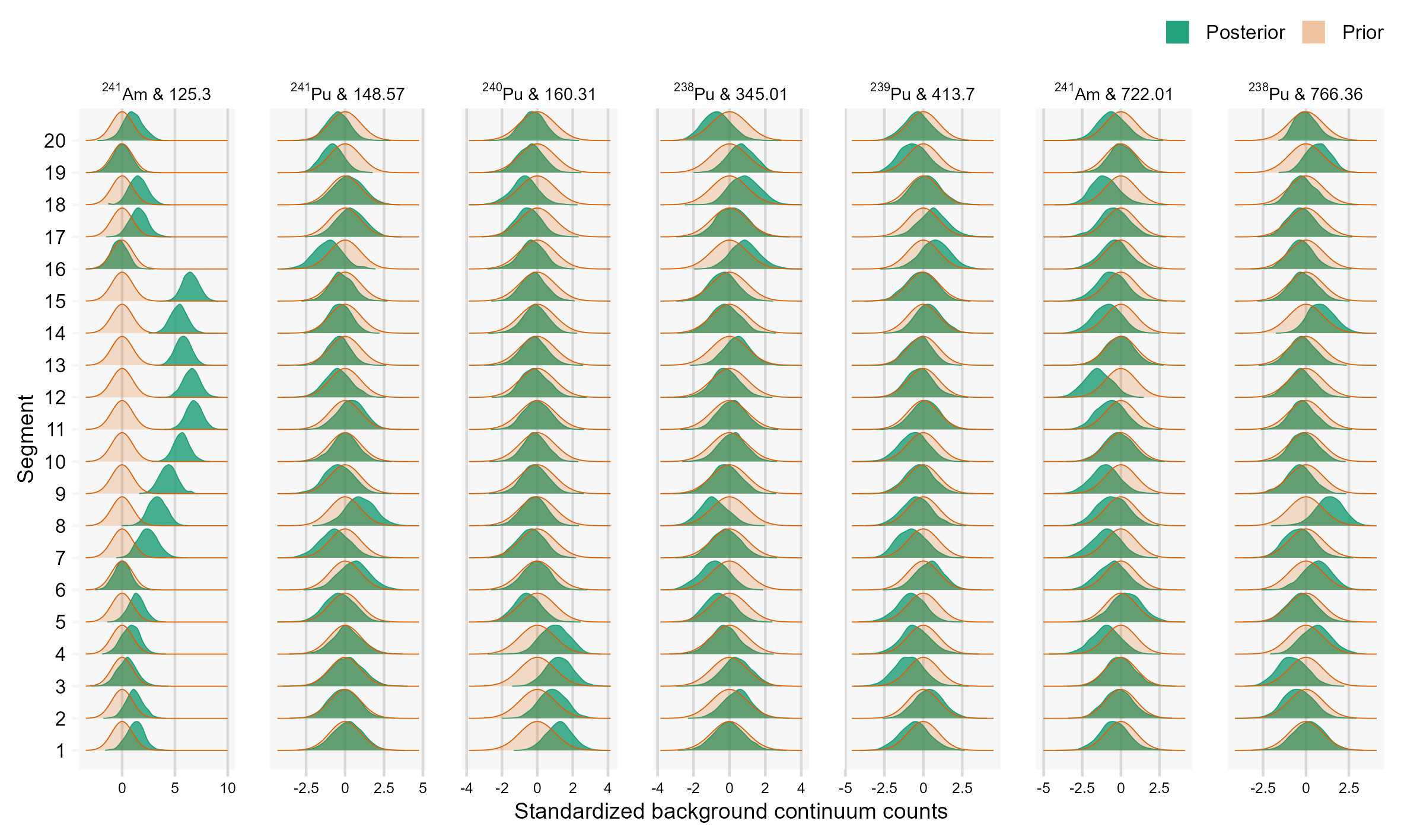}
	\caption{Prior and posterior distributions of the inferred background continuum counts. Each facet's label mentions the considered combination of nuclide and energy peak (keV). For visual convenience, the counts have been standardized using the normal prior parameters and only 7 out of the considered 12 peaks are shown (the following energy peaks are left out of the plot: $^{\rm 241}$Am \& 662.24 keV, $^{\rm 238}$Pu \& 152.72 keV, $^{\rm 239}$Pu \& 129.3 keV, $^{\rm 239}$Pu \& 375.05 keV and $^{\rm 239}$Pu \& 451.48 keV). Segment numbering goes from bottom (1) to top (20).} 
	\label{fig16}
\end{figure}

Going back to our reference strategy 1, Figure \ref{fig16} presents the prior and posterior background count distributions for some selected peaks. Very similar distributions are obtained for strategies 2, 3 and 4 (not shown) while our previous work with the same 3AX-SGS data but a different MCMC inversion setup also produced rather similar posterior count distributions \citep[see Figure 6 in][]{Laloy2021}. As stated by \citet{Laloy2021}, it is worth noting that the prior $p\left(\textbf{b}\right)$ is based on a measured background continuum count that is considered to be a realization of a Poisson distribution of which the shape parameter is unknown. By using $p\left(\textbf{b}\right) = Pois\left(\textbf{b}\right)$ (or $p\left(\textbf{b}\right) = N\left(\textbf{b},\textbf{C}_b\right) \approx Pois\left(\textbf{b}\right)$ as done herein) one implicitly postulates that the measured background continuum count is equal to the mean of its underlying distribution, which is obviously not necessarily the case. Some deviations of $p\left(\textbf{b} | \textbf{d}\right)$ from $p\left(\textbf{b}\right)$ are thus to be expected.

\begin{figure}[hbt!]
	\noindent\hspace{0cm}\includegraphics[width=35pc]{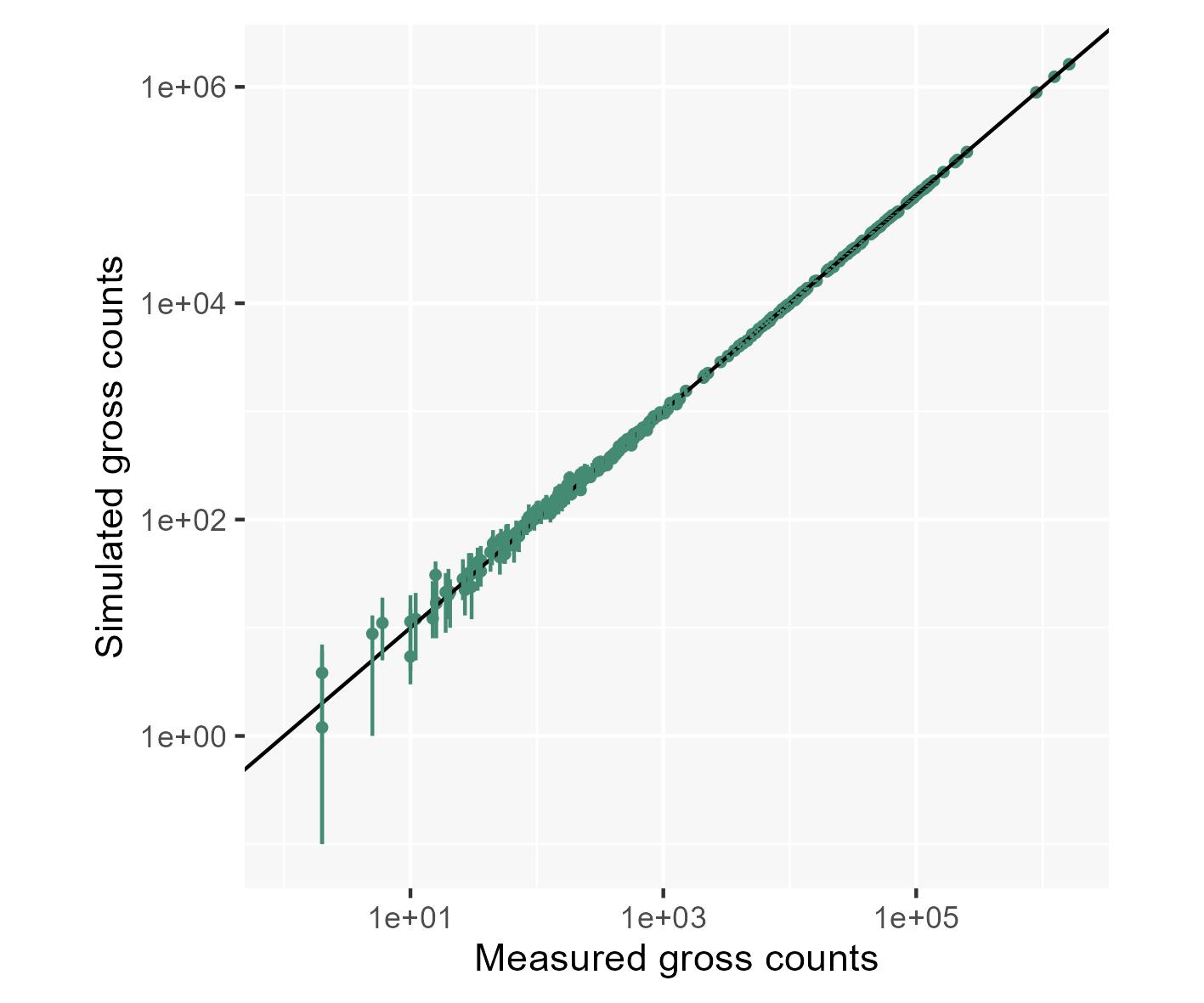}
	\caption{Measured gross counts against a posteriori simulated gross counts. The green points represent the maximum likelihood solution (solution with the largest Poisson likelihood) among the collected 40,000 posterior samples, and the vertical bars denote the 95\% posterior uncertainty intervals. A base 10 logarithmic scale is used to make the deviations between the smallest observed and simulated gross counts visible.} 
	\label{fig17}
\end{figure}

Lastly, Figure \ref{fig17} shows a so-called posterior predictive check where the fit to the 3AX-SGS data provided by the derived posterior solutions is verified. A base 10 logarithmic scale is used to make it possible to visualize the discrepancies between (very) small measured and simulated counts while vertical bars denote the 95\% posterior uncertainty intervals of the simulated counts. This plot is essentially similar to that obtained in our previous study \citep[see Figure 7 in][]{Laloy2021}. The measured gross counts are globally well appproximated as the 95\% uncertainty intervals are relatively tight and most often include the 1:1 line. Moreover, the most important relative deviations are for some pretty small gross counts, below 10 $-$ 15. Overall, these fitting results are consistent with both our expectations and the Poisson count statistics. As of the PNCC measurement, the measured number of reals per second is also very well fitted. The maximum likelihood simulated value is practically equal to the measured value and its associated relative standard deviation is equal to assumed relative standard deviation of 5 \% in the PNCC likelihood (equation \ref{mcmc5}).

\FloatBarrier

\subsection{Computational time}

On the used 6-core workstation and for the used algorithmic parameters of greta/HMC, performing one MCMC iteration in each of the Markov chains takes about 0.5 s (this is an average over warmup and sampling iterations as greta warmup iterations incur a larger computational cost that sampling iterations). Therefore, performing 10,000 warmup iterations and 5000 sampling iterations, as done with our inversion strategies 1 and 2, roughly takes 2 hours. Such runtime per drum might be prohibitively long for routine analyses, let alone that inversion strategy 3 requires twice more warmup iterations that strategies 1 and 2 to reach $\hat{R}$-convergence at the 1.1 level. This warrants further investigations on how to accelerate the inference.

\section{Concluding remarks}
\label{conclusion}

We demonstrate in this study three methodological improvements for Bayesian inference of the radionuclide inventory in radioactive waste drums. First we rely on the Dirichlet distribution for the prior distribution of the isotopic vector as this distribution has the nice property that the elements of its vector samples sum up to 1. Second, we show that such Dirichlet priors can be incorporated within an hierarchical modeling of the prior uncertainty in the isotopic vector, when prior information about isotopic composition is available. Our used Bayesian hierarchical modeling framework makes use of this available information but also acknowledges its uncertainty by letting to a controlled extent the information content of the indirect measurement data (i.e., gamma and neutron counts) shape the actual prior distribution of the isotopic vector(s). Third, we propose to take advantage of GP prior modeling when inferring 1D spatially-distributed mass or, equivalently, activity distributions, to help regularize the MCMC inversion. Combining hierarchical modeling of the prior isotopic composition uncertainty together with GP prior modeling of the vertical Pu profile across the drum appears to work rather well. We find that our GP prior approach can handle both cases with and without spatial correlation, and additionally speeds up the inference, even when no correlation is present, compared to using a log-uniform prior for the Pu content in each considered drum's segment.

Of course, our proposed GP prior modeling framework only makes sense in the context of spatial inference. Furthermore, the computational times involved by our proposed approach are on the order of a few hours, say about 2, to provide uncertainty estimates for all variables of interest. This might prevent our approach to be applied for routine analyses and future work will focus on speeding up the inference. With respect to uncertainty in the efficiencies, our study uses the same stylized drum modeling approach as proposed by \citet{Laloy2021} to account for the source distribution uncertainty across the vertical direction of the drum within the Bayesian inversion. Ongoing investigations on combining different measurement methods within a Bayesian framework that are also performed within the CHANCE project rely on the same approach to handle matrix-related uncertainties. These results will be presented in due course.

\section{Acknowledgments}
\label{acknowledgments}

This work received funding by the EU project CHANCE ``Characterization of Conditioned Nuclear Waste for its Safe Disposal in Europe". An example code of the proposed approach is available at \url{https://github.com/elaloy/UQ-RADWASTE}.

\appendix

%\section{}
%\label{appendix}



\end{document}